\documentclass[preprint,onecolumn,superscriptaddress,amsmath,amssymb,aps,floatfix,nofootinbib,longbibliography]{revtex4-1}

\usepackage[pdftex]{graphicx}
\usepackage{bm}
\usepackage{color}
\usepackage{times}
\usepackage{amsmath}
\usepackage{amssymb}
\usepackage{mathtools}
\usepackage{gensymb} 
\newcommand{\defeq}{\vcentcolon=}
\usepackage{tablefootnote}
\usepackage{textcomp}

\begin{document}

\title{Large poroelasto-plastic deformations \\ due to radially outward fluid injection}

\author{Lucy C. Auton}
\affiliation{Mathematical Institute, University of Oxford, Oxford, OX2 6GG, UK}
\affiliation{Department of Engineering Science, University of Oxford, Oxford, OX1 3PJ, UK}
\author{Christopher W. MacMinn}
\email{christopher.macminn@eng.ox.ac.uk}
\affiliation{Department of Engineering Science, University of Oxford, Oxford, OX1 3PJ, UK}

\date{\today}

\begin{abstract}
Flow-induced failure of granular materials is relevant to a broad range of geomechanical applications. Plasticity, which is the inherent failure mechanism of most granular materials, enables large deformations that can invalidate linearised models. Motivated by fluid injection into a borehole, we develop a steady-state model for the large deformation of a thick-walled, partially-permeable, elastic--perfectly-plastic annulus with a pressurised inner cavity. We account for pre-existing compressive stresses, as would be present in the subsurface, by subtracting a compressed initial state from our solutions to provide the additional disturbance due to fluid injection. We also introduce a simple parameter that allows for a smooth transition from an impermeable material (\textit{i.e.}, subject to mechanical loading at the inner wall) to a fully permeable material (\textit{i.e.}, subject to an internal pore-pressure gradient), which would be relevant to coated boreholes and very-low-permeability materials. We focus on the difference between poroelastic and poroelasto-plastic deformations, the role of kinematic and constitutive nonlinearity, and the transition from impermeable to fully permeable. We find that plasticity can enable much larger deformations while predicting much smaller stresses. The former makes model choice increasingly important in the plastic region, while the elastic region remains insensitive to these changes. We also find that, for a fixed total stress at the inner wall, materials experience larger deformations and generally larger stresses as they transition from impermeable to fully permeable.
\end{abstract}

\maketitle

\section{Introduction}

A clear understanding of the failure of granular materials due to fluid injection has direct relevance to a variety of applications in geomechanics, including borehole stability, cavity expansion, and `fracking' for the recovery of oil or natural gas from shales. These problems involve radially outward flow and deformation surrounding a long cylindrical hole through soil or rock. Most soils and shallow sedimentary rocks have a granular microstructure, and are intrinsically porous, water-saturated, and susceptible to plastic failure. These features suggest that these problems should be approached within the framework of large-deformation poroelasto-plasticity, but essentially all previous work has neglected at least one of these three ingredients (large deformations, permeability, or plasticity) and no previous study has assessed their relative importance.

A borehole is a long cylindrical cavity in the subsurface, from which the original material has been removed by drilling. The walls of the borehole can be supported by a metal casing (with or without perforations), or can be uncased and self-supporting. The stability of a borehole depends on many factors, including the state of stress in the material before drilling, the orientation of the borehole, and the loading to which the borehole is subjected. Problems in borehole integrity are concerned with predicting and preventing borehole collapse after drilling, or during subsequent operations (\emph{e.g}., during fluid injection or extraction). Borehole integrity has been studied extensively \citep[\textit{e.g},][]{wang1996effect, wang1994stresses, detournay1988poroelastic, risnes1982sand, wang1991borehole, wang1991boreholeyield, wang1994boreholerupture, detournay1987two}, but exclusively under the assumption of infinitesimal deformations. \citet{detournay1987two} and \citet{wang1991borehole} neglect the permeability of the material while allowing for plastic failure, whereas \citet{detournay1988poroelastic} consider nonzero permeability while neglecting plasticity.

In cavity expansion, a borehole-like cavity is created via radially-outward mechanical displacement of material (typically soil) around an insertion point. Cavity expansion is a classical problem in geotechnical engineering, with relevance to pile-driving and penetrometer testing \citep{vesic1972expansion, carter1986cavity, yu1991finite, yu2000cavity, davis2002plasticity, howell2009applied}. Large plastic deformations are central to this process; hence, all of these studies consider rigorous large-deformation elasto-plasticity. However, for simplicity, they focus on drained or undrained limiting behaviours and thus neglect the role of fluid flow.

In fracking, fluid is injected into hydrocarbon-bearing rocks, usually shales, in order to open fractures around the injection point. These fractures provide hydraulic access deeper into the reservoir and allow gas to be collected from a larger region of the rock~\citep{pye1973hydraulic,economides2000reservoir, haimson1969hydraulic}. Fracking relies on the brittle failure of shale, but the mechanical properties of shale depend strongly on the composition \citep{economides2000reservoir, britt2009geomechanics, rickman2008practical}; many hydrocarbon-bearing shales have high clay content and may therefore have non-negligible ductility \citep{daigle2014porosity, swift2014nano, vega2014ct,vallejo1988brittle}. Both laboratory and field data suggest that the ductility of shales can have an important impact on the success of fracking~\citep{haimson1969hydraulic, britt2009geomechanics}, but the ductility of shales is almost always neglected. Most studies also neglect fluid flow through the shale due to its extremely low permeability~\citep[\textit{e.g.,}][]{wang1991borehole, wang1994boreholerupture, detournay2004propagation, hubbert1972mechanics}.

Here, motivated by these problems, we develop a kinematically rigorous, steady-state model for the large deformation of a porous, thick-walled, elastic--perfectly-plastic annulus with a pressurised inner cavity. To explore the role of fluid flow (\textit{i.e.}, the importance of permeability), we introduce a new parameter that allows us to transition smoothly from an impermeable material to a fully permeable material. This parameter is physically analogous to the presence of a thin, weak, low-permeability `skin' near the cavity wall. In practise, such regions form naturally in boreholes when drilling and/or injection push fine grains into the pore space, or artificially due to the injection of `wall-building' chemicals. The latter enables a fixed volume of fluid to support the walls of the borehole via hydrostatic pressure. To account for pre-existing stresses, as would be present around a borehole in the subsurface, we consider the impact of fluid injection relative to a pre-stressed initial state. We show that plastic failure enables large deformations that make it essential to account for rigorous, nonlinear kinematics in the plastic region. The elastic region is insensitive to these changes. We also show that, for a given applied total stress at the inner radius, a fully permeable material deforms much more than an impermeable one.

\section{Theoretical Model}

We consider a simple model for the fluid-driven deformation of a thick-walled annulus with a pressurised inner cavity. We take the annulus to be made of a homogeneous cohesive granular material, having relaxed outer radius $b^\mathrm{ref}$ and relaxed inner radius $a^\mathrm{ref}$ (I in Figure~\ref{fig1}). We focus on the final \textit{steady state} during fluid injection; throughout, we assume \textit{axisymmetry}.

\subsection{Pre-stressed initial state}

\begin{turnpage}
\begin{figure}
    \begin{center}
        \includegraphics[width=23cm]{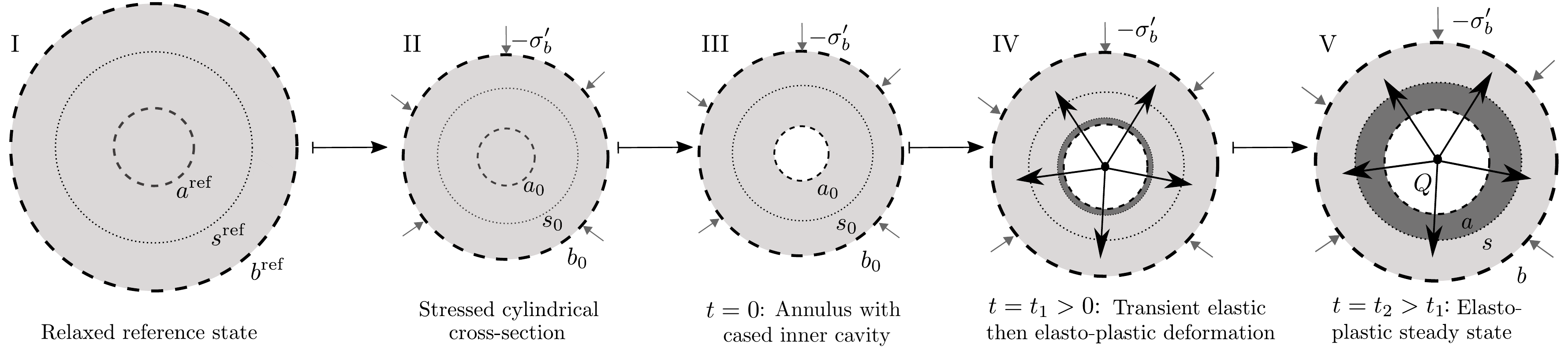}
    \end{center}
    \caption{ Conceptual model for fluid-driven deformation of an annulus with a pressurised inner cavity: (I)~We consider a relaxed, solid cylinder of material---this is our relaxed reference state. (II)~This cylinder is then subjected to a uniform radial compressive effective stress of magnitude $\sigma_b^\prime<0$, as well as a compressive axial effective stress that enforces plane strain; we assume that the cylinder deforms purely elastically. (III)~We then consider the removal of a concentric cylinder of material, followed by the insertion of a permeable ``casing'', such that the stresses and deformation are unchanged---this is our deformed initial state. (IV)~Pressurisation of the cavity leads to elastic and then elasto-plastic expansion of the annulus; we assume that plastic failure initiates at the inner cavity wall and expands outwards, with the inner plastic and outer elastic regions separated by a sharp boundary (dark grey and light grey, respectively). (V)~For steady injection, the flow and deformation will eventually reach a steady state. The inner and outer radii are material surfaces that move over the course of the deformation (dashed lines). The material comprising the inner radius has reference position $a^\mathrm{ref}$, initial position $a_0$, and steady-state position $a$. The material comprising the outer radius has reference position $b^\mathrm{ref}$, initial position $b_0$, and steady-state position $b$. The elastic-plastic interface also evolves, but is not a material surface. The material comprising the elastic-plastic interface at steady state is located at radius $s$; the reference and initial positions of this material are $s^\mathrm{ref}$ and $s_0$, respectively (dotted lines). Note that we take tension to be positive. \label{fig1} }
\end{figure}
\end{turnpage}

The subsurface exists in a state of compressive tectonic stress, the principal values of which typically vary from each other by less than one order of magnitude~\cite{hubbert1972mechanics, bazant2014fracking}. To mimic this compressed initial state, we consider a solid cylinder in plane strain in the plane orthogonal to its axis, with the two in-plane effective stresses equal and given by $\sigma_b^\prime\leq0$. Imposing this stress on the relaxed cylinder decreases its outer radius, $b^\mathrm{ref}\mapsto{}b_0\leq{}b^\mathrm{ref}$ (I${}\mapsto{}$II in Figure~\ref{fig1}); we assume that this compression is purely elastic.

We then assume that a cylinder of material of radius $a_0$ (relaxed radius $a^\mathrm{ref}$) is removed from this compressed state, and that the resulting cavity is supported with a permeable ``casing'' that preserves the radius $a_0$, such that the stresses in the remaining material are unchanged (II${}\mapsto{}$III in Figure~\ref{fig1}). We assume that this casing has no impact on fluid exchange with the surrounding material, or on its outward deformation. Note that there will exist a minimum value of $q$ above which the casing is no longer supporting the cavity.

We then consider the elastic and subsequent elasto-plastic deformation of the resulting annulus due to radially outward fluid injection (III${}\mapsto{}$IV${}\mapsto{}$V in Figure~\ref{fig1}). During injection, both radii expand such that $a_0\mapsto{}a>a_0$ and $b_0\mapsto{}b>b_0$. We assume that plastic yield initiates at the inner boundary and evolves outward to some intermediate radius $s$, such that the material has deformed elasto-plastically for $a\leq{}r<s$ and purely elastically for $s<r\leq{}b$, where $r$ is the radial coordinate. We assume that the effects of injection are localised around the cavity, such that the effective stresses and the pressure tend to their far-field values ($\sigma_b^\prime$ and $0$, respectively) for finite $b$.

We seek the steady-state flow and deformation fields for this problem, for which the values of $a$, $s$, and $b$ are unknown \textit{a priori}. We are interested in the deformation of the material relative to its compressed initial state (III in Figure~\ref{fig1}), but we necessarily derive the steady state relative to the relaxed reference state (IV in Figure~\ref{fig1}) due to the nonlinear nature of the model.

We next present our axisymmetric, plane-strain model for this problem in dimensionless form. We present our scalings and identify key dimensionless parameters in \S\ref{scaling}. We summarise the key aspects of the model, including kinematics, Darcy's law, and mechanical equilibrium, in \S\ref{summary}. We then discuss constitutive laws for the solid skeleton \S\ref{const_laws}.


\subsection{Scaling}\label{scaling}

To write the model in dimensionless form, we adopt characteristic scales for length, stress/pressure, and permeability. We take the dimensional relaxed (reference) outer radius ${b}^\mathrm{ref}$ as the characteristic length scale and the $p$-wave (oedometric) modulus ${\mathcal{M}}$ as the characteristic stress/pressure scale. We model fluid injection as a line source at the origin, characterised by either a fixed dimensionless flow rate $\tilde{q}$ or an applied dimensionless total stress at the inner cavity wall, $\tilde{\sigma}_a$. The dimensionless model is then characterised by the friction angle $\tilde{\varphi}$ and dilation angle $\tilde{\psi}$ (see \S\ref{sec:plastic}), the reference (relaxed) porosity $\tilde{\phi}_f^\mathrm{ref}$, which we take to be uniform for simplicity, and five other dimensionless parameters:
\begin{equation}
    \tilde{\Gamma}\defeq\frac{\Lambda}{\mathcal{M}}, \quad
    \tilde{a}^\mathrm{ref} \defeq\frac{{a}^\mathrm{ref}}{{{b}^\mathrm{ref}}}, \quad
    \tilde{\sigma}_b^\prime \defeq\frac{{\sigma_b}^\prime}{{\mathcal{M}}}, \quad
    \tilde{c} \defeq \frac{{c}}{\mathcal{M}} \quad 
    \text{and either} \quad
    \tilde{q}\defeq\frac{{\mu}{Q}}{2\pi{k}\mathcal{M}} \quad
    \text{or} \quad
    \tilde{\sigma}_a\defeq\frac{{\sigma}_a}{{\mathcal{M}}},
\end{equation}
where ${\Lambda}$ is Lam\'{e}'s first parameter, ${k}$ is the permeability (assumed to be constant and uniform), ${\mu}$ is the dynamic viscosity of the fluid, ${{\sigma}}^\prime_b$ is the radial effective stress (see \S\ref{summary}) at the outer boundary, ${c}$ is the cohesion between grains of the solid skeleton (see \S\ref{sec:plastic}) and ${Q}$ is the volume injection rate per unit length into the page. Note that the dimensionless relaxed outer radius is $\tilde{b}^\mathrm{ref}\equiv1$, and also that we take tension to be positive. Going forward, we work almost entirely in terms of dimensionless quantities and thus drop the tildes ($\tilde{\star}$) for convenience; we denote any further dimensional quantities with a breve ($\breve{\star}$).

\subsection{Kinematics, Darcy's law and mechanical equilibrium}
\label{summary}

Axisymmetry implies that the fluid velocity $\bm{v}_f$ and solid displacement $\bm{u}_s$ each have only one nontrivial component, such that $\bm{v}_f = v_{f}(r)\bm{\hat{e}}_r$ and $\bm{u}_s = u_{s}(r)\bm{\hat{e}}_r$, where $r$ is the Eulerian radial coordinate and $\bm{\hat{e}}_r$ is the radial unit vector. Steady state implies that $v_f$ and $u_s$ are independent of time. We work in an Eulerian (spatial) reference frame, such that the displacement is given by
\begin{equation}
    u_{s} (r)= r-R(r),
\end{equation}
where the Lagrangian radial coordinate $R(r)$ denotes the original position of the material that is at position $r$ in the deformed state.

We assume that both the individual solid grains and the fluid are incompressible, such that deformation occurs through rearrangement of the grains and corresponding changes in the local porosity or void fraction $\phi_f$. We relate $u_s$ to $\phi_f$ via \citep{auton2017arteries}
\begin{equation}\label{eq:phi_to_u}
    \frac{\phi_f-\phi_f^\mathrm{ref}}{1-\phi_f^\mathrm{ref}} =\frac{1}{r}\frac{\mathrm{d}}{\mathrm{d}r}\left(ru_s-\frac{1}{2}u_s^2\right),
\end{equation}
which linearises under the assumption of infinitesimal strains to 
\begin{equation}\label{eq:phi_to_ulin}
    \frac{\phi_f-\phi_f^\mathrm{ref}}{1-\phi_f^\mathrm{ref}} \approx\frac{1}{r}\frac{\mathrm{d}}{\mathrm{d}r}\left(ru_s\right).
\end{equation}

Additionally, mechanical equilibrium requires that 
$\nabla\cdot\bm{\sigma} = 0$, where $\bm{\sigma}$ is the total Cauchy stress and where we have neglected inertia and the effect of gravity. The total stress can be decomposed as $\bm{\sigma}=\bm{\sigma}^\prime-p\bm{I}$, where Terzaghi's effective stress $\bm{\sigma}^\prime$ is the portion of the total stress supported by deformation of the solid and $p$ is the fluid pressure. Combining these ideas for axisymmetric flow and deformation leads to
\begin{equation}\label{govern}
    \frac{\mathrm{d}\sigma^\prime_r}{\mathrm{d} r}+\frac{\sigma^\prime_r-\sigma^\prime_{\theta}}{r}=\frac{\mathrm{d}p}{\mathrm{d} r},
\end{equation}
where $\sigma^\prime_r$ and $\sigma^\prime_{\theta}$ are the radial and azimuthal (hoop) components of the effective stress, respectively. At steady state, conservation of mass for the fluid leads to 
\begin{equation}\label{COM}
    \frac{1}{r}\frac{\mathrm{d}}{\mathrm{d}r} \left(r\phi_fv_f\right) = 0 \quad\implies\quad \phi_fv_f = \frac{q}{r},
\end{equation}
by the definition of a line source of strength $q$.

Finally, we assume that the fluid flows through the solid according to Darcy's law. For simplicity, we assume that the permeability remains constant. For steady axisymmetric flow, this leads to
\begin{equation}\label{Darcy}
    \phi_fv_f = -\frac{\mathrm{d}p}{\mathrm{d}r} \quad\implies\quad \frac{\mathrm{d}p}{\mathrm{d}r}=-\frac{q}{r}.
\end{equation}
Combining Equations~(\ref{COM}) and (\ref{Darcy}) and integrating leads to a direct relationship between $q$ and the pressure drop $\Delta p$:
\begin{equation}\label{q_dp}
    \Delta{p}\defeq p(a)-p(b) = q \ln\left(\frac{b}{a}\right).
\end{equation}
For a more detailed derivation and discussion of the above aspects of the model, see \citet{auton2017arteries, auton2018arteries}.

\subsection{Elasticity and plasticity}\label{const_laws}

\subsubsection{Elasticity}

Elastic deformations are quasi-static and reversible. Most ductile materials will yield before experiencing moderate or large elastic deformations, so we restrict our attention to infinitesimal deformations and therefore linear elasticity in the elastic region. In linear elasticity, the effective stress is related to the strain via
\begin{subequations}\label{lin_elast}
    \begin{equation}\label{linconstsigr}
        \sigma^\prime_r=\varepsilon_r+\Gamma\varepsilon_\theta, \quad   \sigma^\prime_\theta=\Gamma\varepsilon_r+ \varepsilon_\theta,  \quad \mathrm{and} \quad \sigma_z^\prime = \Gamma( \varepsilon_r +\varepsilon_\theta) = \frac{\Gamma(\sigma_r^\prime+\sigma_\theta^\prime) 
}{1+\Gamma},
\end{equation} 
and the strain is related to the displacement via 
\begin{equation}
\label{elasticstrainthetand}
\varepsilon_r = \frac{\mathrm{d} u_{s}}{\mathrm{d} r}, \quad \varepsilon_\theta = \frac{u_{s}}{r}, \quad \mathrm{and} \quad \varepsilon_z \equiv 0,
\end{equation}
\end{subequations}
where $\varepsilon_r$, $\varepsilon_\theta$, and $\varepsilon_z$ are the radial, azimuthal, and axial components of the strain, respectively, and the expressions for $\sigma_z^\prime$ and $\varepsilon_z$ are consequences of plane strain.


\subsubsection{Plasticity}\label{sec:plastic}

Plastic deformations are path-dependent and irreversible. Plastic failure implies that, once the state of stress exceeds a threshold, some fraction of any additional strain energy will be dissipated through irreversible rearrangements as opposed to being stored elastically. The simplest form of plasticity is `perfect plasticity', in which the material properties are assumed to remain constant after yield and all additional strain energy is dissipated \cite{hill1950mathematical}. The threshold that defines the transition from elastic to plastic behaviour is known as a yield condition. For granular materials, this is typically based on the idea of internal Coulomb-like friction. This implies that the effective shear stress $\tau^\prime$ anywhere within the material must be strictly less than a specified fraction of the corresponding effective normal stress $\sigma^\prime$ for the deformation to remain elastic. For perfect plasticity, $\tau^\prime$ is enforced to remain equal to this fraction of $\sigma^\prime$ after yield. For a cohesive granular material, it is standard to additionally include a yield strength $c$ due to the cohesion between the grains. The simplest cohesive-frictional yield condition is the cohesive Mohr-Coulomb condition, $|\tau^\prime|\leq -\sigma^\prime\tan\varphi+c$,
where $\varphi$ is the friction angle \cite{davis2002plasticity}. We then express this condition in terms of the principal effective stresses\footnote{The principal effective stresses are the eigenvalues of the effective stress tensor, with $\sigma^\prime_1$ and $\sigma^\prime_3$ being largest (most tensile) and the smallest (least tensile), respectively.} $\sigma_1^\prime\geq\sigma_2^\prime\geq\sigma_3^\prime$ by introducing a yield function $\mathcal{F}_{1,3}$,
\begin{equation}
    \mathcal{F}_{1,3}\defeq \alpha\sigma_1^\prime-\sigma_3^\prime -y,
\end{equation}
where
\begin{equation}
    \alpha\defeq\frac{1+\sin(\varphi)}{1-\sin(\varphi)} \quad \text{and}\quad y\defeq\frac{2c\cos(\varphi)}{1-\sin(\varphi)}. 
\end{equation}
Plastic yield now corresponds to the condition $\mathcal{F}_{1,3}=0$. For $\mathcal{F}_{1,3}<0$, the material remains elastic. As soon as equality is first achieved, we enforce $\mathcal{F}_{1,3}\equiv0$ locally thereafter. Taking $\varphi=0$ ($\alpha=1$) provides the Tresca yield condition, which is commonly used to model cohesive soils during \textit{undrained} deformation (\textit{i.e.}, at fixed pore volume) because the associated plastic flow law is volume conservative (see below). The simplicity of the Tresca model enables analytical solutions in many cases~\citep{auton-thesis-2018}. Additionally, these solutions can be used to derive the corresponding solutions for a von Mises material. The von Mises yield condition is commonly used to model multi-axial loading in metals.

Here, the three principal stresses are $\sigma^\prime_r$, $\sigma^\prime_\theta$, and $ \sigma^\prime_z$, but not necessarily in this order. To apply the correct yield condition, it is necessary to determine the correct ordering of these stresses. For a cylinder in plane strain, it is commonly assumed that $\sigma^\prime_1=\sigma^\prime_\theta$ and $\sigma^\prime_3=\sigma^\prime_r$ (\textit{i.e.}, $\sigma_\theta^\prime>\sigma_z^\prime>\sigma_r^\prime$) \cite{wang1991borehole, wang1991boreholeyield}, such that the appropriate yield function is
\begin{equation}\label{yieldcrit}
    \mathcal{F}_{\theta, r}\defeq\alpha{\sigma^\prime}_\theta-{\sigma^\prime}_r - y.
\end{equation}
This ordering has been justified heuristically for an impermeable cylinder~\cite{yu1991finite}. We show in Appendix~\ref{which_yield} that Equation~\eqref{yieldcrit} is indeed the appropriate yield function if the cylinder is sufficiently ``strong''---that is, if $y(b^2-a^2)>-2b^2\sigma_b^\prime$ and $\alpha>(1+\Gamma)/\Gamma$, assuming that $\mathcal{F}_{1,3}$ is maximised at $r=a$ and at steady state, and that, once the material yields according to a given yield condition, the material subsequently yields exclusively according to that condition. These assumptions are consistent with previous work \cite{wang1991boreholeyield, wang1991borehole, yu1991finite}.

We assume that the strain everywhere can be additively decomposed into elastic and plastic components \cite{davis2002plasticity},
\begin{equation}\label{adddecomp}
    \varepsilon = \varepsilon^\mathrm{e}+\varepsilon^\mathrm{p},
\end{equation}
where the superscripts $\mathrm{e}$ and $\mathrm{p}$ denote the elastic and plastic components of the strain, respectively. 
Stored elastic energy is associated with the elastic component of the strain, which is related directly to the effective stress via the elasticity law. Dissipation is associated with the plastic component of the strain, which has no direct connection to the effective stress; instead, the evolution of the plastic strain is described by a plastic flow law. Note that, whereas the strain decomposes into elastic and plastic components, there is one unique stress field.

For a Mohr-Coulomb yield condition, the so-called non-associated flow law is given by
\begin{equation}\label{flow}
    \beta\dot{\varepsilon}^\mathrm{p}_r  +\dot{\varepsilon}^\mathrm{p}_\theta = 0,
\end{equation}
where the overdots denote the material time derivative following the solid and
\begin{equation}
    \beta\defeq\frac{1+\sin(\psi)}{1-\sin(\psi)}.
\end{equation}
A nonzero dilation angle $\psi$ incorporates the fact that most granular materials undergo volumetric expansion (dilation) during plastic flow. For $\psi=\varphi$ ($\beta=\alpha$), this flow law is said to be `associated'; however, associated flow typically overestimates dilation for granular materials \cite{yu2000cavity, salgado2010stress}. For $\psi=0$, this flow law reduces to the associated flow law for the Tresca yield condition and becomes volume conservative (no dilation). For the scenarios considered here, we expect that $0\leq\psi\leq\varphi$ ($1\leq\beta\leq\alpha$).

\subsection{Boundary conditions}\label{theoret}

\subsubsection{Outer boundary}

\begin{subequations} 
At the outer boundary, we impose a constant radial effective stress equal to the confining stress,
\begin{equation}\label{outerbc}
    \sigma_r^\prime(b) = \sigma_b^\prime.
\end{equation}
This means that the outer boundary is a free boundary, at which the relevant kinematic condition is
\begin{equation}\label{kob}
    u_s(b) = b-1.
\end{equation}
Lastly, we take $p(b)=0$.

Note that we assume that $a\leq s< b$, so that the cylinder is not entirely plastic and hence the outer boundary conditions will always be applied to the elastic region. Note that the problem becomes over-constrained, and thus ill-defined, once $s$ reaches $b$.
\end{subequations}

\subsubsection{The elastic-plastic interface}
\begin{subequations}

At the elastic-plastic interface, we enforce continuity of displacement and of radial effective stress,
\begin{equation}\label{mix}
    u_{s}(s^+) =u_{s}(s^-) \quad \mathrm{and} \quad \sigma^\prime_r(s^+) =\sigma^\prime_r(s^-),
\end{equation}
where $s^-$ denotes $r=s$ as approached from within the plastic region $(a\leq r<s)$ and, likewise, $s^+$ denotes $r=s$ as approached from within the elastic region $(s<r\leq b)$. In addition, we require that the effective stresses in the elastic region must be at the point of yield at $r=s^+$,
\begin{equation}\label{atyield}
    \alpha\sigma^\prime_\theta(s^+)-\sigma^\prime_r(s^+)=y.
\end{equation}
\end{subequations}
Note that Equations~\eqref{mix} and \eqref{atyield}, and the fact that $\mathcal{F}_{\theta,r}\equiv0$ throughout the plastic region, together imply continuity of $\sigma_\theta^\prime$ across $r=s$.

\subsubsection{Inner boundary}\label{s:inner_bc}

We suppose that the cavity wall is coated by a thin, weak, low-permeability skin; for a borehole, this skin could be caused by local damage, clogging, or wall-building chemicals. We then suppose that the cavity is pressurised to a pressure $-\sigma_a$. The presence of the skin leads to a partitioning of this pressure between the fluid and the solid, such that the fluid pressure drops by some amount across the skin and the skin then exerts an effective stress on the solid at the cavity wall. For an impermeable skin, this would lead to an imposed radial effective stress $\sigma^\prime_r(a)=\sigma_a$; we refer to this as an `impermeable material'. For an `infinitely' permeable skin, this would lead to an imposed fluid pressure $\Delta{p}\equiv{}p(a)=-\sigma_a$; we refer to this as a `fully permeable material'. To transition smoothly between these two limiting cases, we introduce a new `permeability-load parameter' $\zeta$. We define $\zeta\in[0,1]$ such that the material can vary continuously from impermeable ($\zeta \equiv 0$) to fully permeable ($\zeta \equiv1$). Pressurisation of the cavity then leads to two conditions at the inner boundary,
\begin{subequations}
\begin{equation}\label{innerbcA}
    \sigma^\prime_r(a)= (1-\zeta)\sigma_a
\end{equation}
and 
\begin{equation}\label{dp_sig}
    p(a) \equiv \Delta{p} = -\zeta\sigma_a,
\end{equation}
such that $\sigma_r(a)=\sigma^\prime_r(a)-p(a)\equiv\sigma_a$ for all $\zeta$.

We drive the system by imposing either $\sigma_a$ or the total flow rate $q$. The latter appears explicitly in Equation~\eqref{Darcy}. We express $q$ in terms of $\sigma_a$ using Equations~\eqref{q_dp} and \eqref{dp_sig},
\begin{equation}\label{q_sub}
    q = -\frac{\zeta\sigma_a}{\ln\left(\frac{b}{a}\right)}.
\end{equation}
Note that, for injection, we expect $\sigma_a<0$. For fully permeable materials ($\zeta\equiv{}1$), in particular, it is convenient to impose $q$ rather than $\sigma_a$ (\S\ref{s:figs_q}).

The inner boundary is also a material boundary. As such, the displacement must also satisfy a kinematic condition given by
\begin{equation}\label{kib}
    u_{s}(a)=a-a^\mathrm{ref},
\end{equation}
where $a^\mathrm{ref}$ is the relaxed reference position of the material that eventually comprises the cavity wall.
\end{subequations}

\section{Governing equations and model summary}

\subsection{Relaxed reference state to initial state}\label{s:ref_to_0}

We now consider the transition from the relaxed reference state (I, Figure~\ref{fig1}) to the initial state prior to pressurisation (III, Figure~\ref{fig1}), which we take to be purely elastic compression under an imposed radial effective stress $\sigma^\prime_b$ in plane strain. As with $a$ and $b$, we denote quantities in the initial state by a subscript `0'.

Using Equations~(\ref{lin_elast}) to eliminate the stresses from Equation~(\ref{govern}) in favour of the displacement, and further setting $q\equiv{}0$, purely mechanical linear-elastic deformation is governed by
\begin{equation}
    \frac{\mathrm{d}^2u_{s,0}}{\mathrm{d}r^2} +\frac{1}{r}\frac{\mathrm{d}u_{s,0}}{\mathrm{d}r}-\frac{u_{s,0}}{r^2}= 0
\end{equation}
subject to 
\begin{equation}
    \sigma^\prime_{r,0}(b) = \sigma_b^\prime
\end{equation}
and the requirement of boundedness at the origin. This problem has solution 
\begin{equation}\label{u}
    u_{s,0}(r) = \left(\frac{\sigma_b^\prime}{1+\Gamma}\right)r,
\end{equation}
which gives $\sigma_{r,0} = \sigma_{\theta,0} = \sigma_b^\prime$, $\sigma_{z,0}=2\Gamma\sigma^\prime_b/(1+\Gamma)$, and $p_0 = 0$.

For a rigorous treatment of the kinematics, we impose the outer boundary conditions at $b_0$ rather than at $b^\mathrm{ref}$, and we calculate the initial porosity field according to Equation~\eqref{eq:phi_to_u}. We denote this initial state with a superscript `Q' to indicate that it combines linear elasticity with rigorous kinematics (`Quasi-linear'). The `Q' initial porosity is given by
\begin{equation}\label{phiQ}
    \phi_{f,0}^Q= \phi_f^\mathrm{ref}+\frac{\sigma_b^\prime(1-\phi_f^\mathrm{ref})[2(1+\Gamma)-\sigma_b^\prime]}{(1+\Gamma)^2}.
\end{equation}
We eliminate the Eulerian coordinate $r$ from Equation~(\ref{u}) in favour of the Lagrangian coordinate $R$ via the kinematic relationship $u_{s,0}(r)=r-R$,
\begin{equation}
    u_{s,0}^Q =\left(\frac{\sigma_b^\prime}{1+\Gamma-\sigma_b^\prime}\right)R.
\end{equation} 
Finally, Equation~\eqref{kib} leads to
\begin{equation}\label{arefQ}
    a_0^Q =\left(\frac{1+\Gamma}{1+\Gamma-\sigma_b^\prime}\right)a^\mathrm{ref}.
\end{equation}

We also derive a fully linearised version of the initial state by imposing the outer boundary conditions at $b^\mathrm{ref}\equiv{}1$ and by calculating the initial porosity field according to Equation~\eqref{eq:phi_to_ulin}. We denote this initial state with a superscript `L' to indicate that it is fully linearised under the assumption of infinitesimal strains (`Linear'). The `L' initial state is given by
\begin{equation}
    \phi_{f,0}^L = \phi_f^\mathrm{ref}+\frac{2(1-\phi_f^\mathrm{ref})\sigma_b^\prime}{1+\Gamma}, \quad u_{s,0}^L(R) = \left(\frac{\sigma_b^\prime}{1+\Gamma}\right)R, \quad \text{and}\quad a_{0,L} =\left(\frac{\sigma_b^\prime+1+\Gamma}{1+\Gamma}\right)a^\mathrm{ref}.
\end{equation}
Note that Equation~\eqref{kib} is not needed in this case.

\subsection{Initial state to steady state}

We now seek solutions to the poroelasto-plastic problem at steady state (V in Figure~\ref{fig1}). We denote the displacement field in this state by $u_s$.

\subsubsection{Elastic region}

We again use Equations~(\ref{lin_elast}) to eliminate the stresses in Equation~(\ref{govern}) in favour of the displacement, and we now also use Equation~(\ref{q_sub}) to eliminate $q$ in favour of $\sigma_a$. This leads to an ordinary differential equation (ODE) in $u_s$ in the elastic region,
\begin{equation}\label{a_p}
    \frac{\mathrm{d}^2u_s}{\mathrm{d} r^2} + \frac{1}{r}\frac{\mathrm{d}u_s}{\mathrm{d}r}-\frac{u_s}{r^2}=\frac{\zeta\sigma_a}{r\ln\left(\frac{b}{a}\right)} \quad s<r<b. 
\end{equation}
The domain of this second-order ODE has two free boundaries---a material boundary at $r=b$ and a constitutive boundary at $r=s$---and the solution will involve two constants of integration. We again define a `Q' class of models subject to rigorous kinematics, where porosity is calculated according to Equation~\eqref{eq:phi_to_u} and which requires four boundary conditions:
\begin{subequations}\label{BCelastQ}
\begin{align}
    &\alpha\sigma_\theta^\prime(s^+)-\sigma_r^\prime(s^+)=y,  \label{BCelastQa} \\
    & u_s(s^+) = u_s(s^-), \label{BCelastQb} \\
    &\sigma_r^\prime(b)  = \sigma_b^\prime,  \label{BCelastQc}
\end{align}
and
\begin{equation}\label{BCelastQd}
    \hspace{-1.6cm} u_s(b) = b-1. 
\end{equation}
\end{subequations}
We also again define an `L' class of models that are fully linearised under the assumption of infinitesimal strains, in which we calculate the porosity according to Equation~\eqref{eq:phi_to_ulin} and apply Equation~\eqref{BCelastQc} at $b^\mathrm{ref}\equiv{}1$. Equation~\eqref{BCelastQd} is not needed in this case.

\subsubsection{Plastic region}

In the plastic region, combining Equation~\eqref{yieldcrit} with Equation~\eqref{govern} and using Equation~(\ref{q_sub}) to eliminate $q$ in favour of $\sigma_a$ leads to
\begin{equation}\label{comboepA}
    \frac{\mathrm{d}\sigma^\prime_r}{\mathrm{d}r} +\frac{\sigma_r'(\alpha-1)}{\alpha r} =\frac{y\ln\left(\frac{b}{a}\right)+\alpha \zeta\sigma_a}{\alpha r\ln\left(\frac{b}{a}\right)} \qquad a<r<s,
\end{equation}
which has solution
\begin{equation}\label{plasticstressA}
    \sigma_r^\prime = C_1+C_2\left(\frac{r}{a}\right)^K \quad\implies\quad \sigma_\theta^\prime = \frac{y+\sigma_r^\prime}{\alpha} = \frac{1}{\alpha}\left[y+C_1 +C_2\left(\frac{r}{a}\right)^K\right],
\end{equation}
where 
\begin{equation}\label{defsAA}
    K  \defeq \frac{1-\alpha}{\alpha}, \quad
C_1(a,b) \defeq \frac{y\ln\left(\frac{b}{a}\right)+\alpha\sigma_a\zeta}{(\alpha-1)\ln\left(\frac{b}{a}\right)}, \quad\text{and}\quad
C_2(a,b) \defeq  -C_1+(1-\zeta)\sigma_a.
\end{equation}
To determine the displacement, we integrate Equation~\eqref{flow} with respect to time to arrive at
\begin{equation}\label{flow2}
    \beta\varepsilon_r^\mathrm{p} +\varepsilon_\theta^\mathrm{p}= f,
\end{equation}
where $f$ is a quantity whose material time derivative following the solid must vanish. Equation~(\ref{flow}) is a statement that the quantity $\beta\varepsilon_r^\mathrm{p} +\varepsilon_\theta^\mathrm{p}=f$ must be conserved (\textit{i.e.}, $f$ is a conservative tracer advected with the solid). Prior to fluid injection, there is no plastic strain in any direction ($\varepsilon_r^\mathrm{p}=\varepsilon_\theta^\mathrm{p}=0$) and therefore $f=0$; hence, it must be the case that $f\equiv0$. We then decompose the plastic strains according to Equation~\eqref{adddecomp},
\begin{equation}\label{flowsimpA}
    \beta\varepsilon_r +\varepsilon_\theta =\beta\varepsilon^\mathrm{e}_r +\varepsilon^\mathrm{e}_\theta \qquad a<r<s.
\end{equation}
The elastic strains are always related to the effective stresses via the elasticity law. As such, Equations~\eqref{linconstsigr} and \eqref{plasticstressA} allow us to rewrite the right-hand side of Equation~\eqref{flowsimpA} as
\begin{subequations}
\begin{equation}\label{41aA}
    \beta\varepsilon^\mathrm{e}_r +\varepsilon^\mathrm{e}_\theta =\frac{\sigma^\prime_r(\beta-\Gamma)+\sigma^\prime_\theta(1-\beta\Gamma)}{1-\Gamma^2}=D_1+D_2\left(\frac{r}{a}\right)^K,
\end{equation}
where
\begin{equation}\label{defsA}
D_1(a,b)\defeq \frac{\alpha C_1(\beta-\Gamma)+(y+C_1)(1-\beta\Gamma)}{\alpha(1-\Gamma^2)}
\end{equation}
and
\begin{equation}
    D_2(a,b)\defeq\frac{C_2\left[\alpha(\beta-\Gamma) +1+\beta\Gamma\right]}{\alpha(1-\Gamma^2)} \equiv \frac{C_2}{C_1}\left[D_1-\frac{y(1-\beta\Gamma)}{\alpha(1-\Gamma^2)}\right], 
\end{equation}
\end{subequations}
and $C_1$ and $C_2$ are as defined in Equations~\eqref{defsAA}.

The plastic strains can, in general, be large. Hence, it may be appropriate to adopt a nonlinear constitutive model for the total strains in the plastic region. Here, we consider the Hencky (logarithmic) model \cite{yu2000cavity, bazant1998easy}. For Hencky strains, the left-hand side of Equation~\eqref{flowsimpA} becomes
\begin{equation}\label{flowhenkA}
    \beta\varepsilon_r+\varepsilon_\theta = -\ln\left[\left({1-\frac{\mathrm{d}u_s}{\mathrm{d}r}}\right)^\beta\left({1-\frac{u_s}{r}}\right)\right],
\end{equation}
which, on combining Equation~\eqref{flowhenkA} with Equations~\eqref{flowsimpA} and \eqref{41aA}, leads to
\begin{equation}\label{finalA}
    \frac{\mathrm{d}u_s}{\mathrm{d}r}= 1-\left(1-\frac{u_s}{r}\right)^{-\frac{1}{\beta}}\exp\left(-\frac{D_1a^K+D_2r^K}{a^K\beta}\right), 
\end{equation}
where $D_1$ and $D_2$ are as defined in Equation~\eqref{defsA}. We denote this nonlinear model by `N' (`Nonlinear').

For comparison, we also consider a linear constitutive model for the total strains in the plastic region,
\begin{equation}\label{linflow}
    \beta\varepsilon_r+\varepsilon_\theta = \beta\frac{\mathrm{d}u_s}{\mathrm{d}r}+\frac{u_s}{r},
\end{equation}
which linearises to Equation~\eqref{linflow} for infinitesimal strains. Combining Equation~\eqref{linflow} with Equations~\eqref{flowsimpA} and \eqref{41aA} yields
\begin{equation}\label{plast_lin}
    \beta \frac{\mathrm{d}u_s}{\mathrm{d}r} +\frac{u_s}{r} = D_1+D_2\left(\frac{r}{a}\right)^K. 
\end{equation}
We denote this model by `Q'. The `Q' model can be further linearised by taking $a\approx{}a^\mathrm{ref}$; we denote this fully linearised model by `L'.

\subsubsection{Model summary}\label{Eqn_sum_NQL}

For rigorous kinematics throughout and Hencky strains in the plastic region, the full boundary value problem (BVP) is 
\begin{subequations}\label{govern21A}
    \begin{align}
        &\frac{\mathrm{d}u_s}{\mathrm{d}r}=  1-\left(1-\frac{u_s}{r}\right)^{-\frac{1}{\beta}}\exp\left(-\frac{D_1a^K+D_2r^K}{a^K\beta}\right)   \qquad a<r<s  \label{plasticA},\\
        &\frac{\mathrm{d}^2u_s}{\mathrm{d} r^2} + \frac{1}{r}\frac{\mathrm{d}u_s}{\mathrm{d}r}-\frac{u_s}{r^2}=\frac{\zeta\sigma_a}{r\ln\left(\frac{b}{a}\right)},  \quad \hspace{3.5cm} s<r<b, \label{elasticA} \\
        & u_s(s^+) = u_s(s^-), \label{cont_u} \\
        & \sigma_r^\prime(s^+) = \sigma_r^\prime(s^-), \label{another_label} \\
        & \alpha\sigma_\theta^\prime(s^+)-\sigma_r^\prime(s^+)=y,  \label{yield_in_elastic}\\
        & \sigma_r^\prime(b) = \sigma_b^\prime, \label{outer_BC} \\
        & u_s(a) = a-a^\mathrm{ref},  \label{kin_a} \\
        & u_s(b) = b-1, \label{kin_b}
    \end{align}
\end{subequations}
where $\sigma_r^\prime(s^+)$, $\sigma_\theta^\prime(s^+)$, and $\sigma_r^\prime(b)$ can be expressed in terms of $u_s$ via Equations~\eqref{lin_elast}, and $\sigma_r^\prime(s^-)$ is given in Equation~\eqref{plasticstressA}.

We denote this model by `NQ', where the first letter refers to rigorous kinematics and nonlinear strains in the plastic region (`Nonlinear') and the second refers to rigorous kinematics and linear elasticity in the elastic region (`Quasi-linear'). Note that Equation~\eqref{innerbcA} has already been imposed in the above. For comparison, we also consider linear strains throughout the entire domain by coupling Equation~\eqref{plast_lin} with Equations~\eqref{elasticA}---\eqref{kin_b}. We denote this model `QQ' to indicate rigorous kinematics and linear strains throughout the domain.

Although deformations and strains in the plastic region may be large, we expect deformations and strains in the elastic region to remain small. Thus, we further linearise the kinematics in the elastic region under the assumption of small deformations to provide an intermediate model defined by Equation~\eqref{plast_lin} and Equations~\eqref{elasticA}---\eqref{kin_a}, with Equation~\eqref{outer_BC} applied at $b\approx{}b^\mathrm{ref}\equiv{}1$. We denote this model by `QL', where the `L' refers to a fully linearised model in the elastic region. This problem involves only two free boundaries (at $a$ and $s$).

Finally, we consider a fully linearised model, which we denote `LL'. This model comprises Equation~\eqref{plast_lin} with $a\approx{}a^\mathrm{ref}$, Equations~\eqref{elasticA}---\eqref{yield_in_elastic}, and Equation~\eqref{outer_BC} evaluated at $b\approx{}b^\mathrm{ref}\equiv{}1$. This problem involves only one free boundary (at $s$) and can be solved analytically, with $s$ determined via an implicit relation that is straightforward to solve numerically using conventional root-finding techniques. This solution provides a good first guess for the numerical solution of the other three models (see \S\ref{numerics}).

\section{Solutions for imposed $\sigma_a$ and general $\zeta$}
\label{s:solutions}

For a given set of parameters, our poroelasto-plastic models will not be well posed for all values of $\sigma_a$. Specifically, the material will remain elastic if $\sigma_a$ is sufficiently small; this purely poroelastic problem was studied extensively by \citet{auton2017arteries, auton2018arteries} for fully permeable materials ($\zeta\equiv1$). Conversely, the material will yield completely if $\sigma_a$ is sufficiently large. We denote the value of $\sigma_a$ at which the material will first yield by $\sigma_{a}^\mathrm{min}$ and the value of $\sigma_a$ at which the material will yield completely by $\sigma_{a}^\mathrm{max}$. Note that $\sigma_{a}^\mathrm{max}<\sigma_{a}^\mathrm{min}<0$.

We next consider $\sigma_{a}^\mathrm{min}$ and $\sigma_{a}^\mathrm{max}$ for all models (\S\ref{minsig} and \S\ref{maxsig}, respectively). We then derive an implicit solution to the LL model (\S\ref{LLk0}). Finally, we discuss the numerical scheme used here for solving the QL, QQ, and NQ models (\S\ref{numerics}).

\subsection{Value of $\sigma_a$ at which yield first occurs ($\sigma_{a}^\mathrm{min}$)}\label{minsig}

We determine $\sigma_{a}^\mathrm{min}$ by considering the purely poroelastic problem. Solving Equation~(\ref{a_p}), we obtain the general expression for the linear elastic displacement,
\begin{subequations}\label{genundzeta}
\begin{equation}
    u_{s}=\frac{\sigma_a\zeta r\ln(r) }{2\ln\left(\frac{b}{a}\right)}+\frac{\mathcal{B}_1r}{1+\Gamma}+\frac{\mathcal{B}_2}{(1-\Gamma)r} - \frac{\zeta\sigma_ar}{2(1+\Gamma)\ln\left(\frac{b}{a}\right)},
\end{equation}
where $\mathcal{B}_1$ and $\mathcal{B}_2$ are unknown functions of $a$ and $b$. The general form of the stresses in the elastic region is then
\begin{align}
    \sigma^\prime_r&= \frac{(1+\Gamma)\zeta\sigma_a\ln(r)}{2\ln\left(\frac{b}{a}\right)} +\mathcal{B}_1-\frac{\mathcal{B}_2}{r^2}, \label{radstresA} \\
    \sigma^\prime_\theta&= \sigma_r^\prime + \frac{2\mathcal{B}_2}{r^2} - \frac{\zeta\sigma_a(1-\Gamma)}{2\ln\left(\frac{b}{a}\right)}. \label{thetastres}
\end{align}
\end{subequations}
For the poroelastic problem, we then apply the boundary conditions $\sigma_r^\prime(a) = (1-\zeta)\sigma_a$ and $\sigma_r^\prime(b)= \sigma_b^\prime$ to Equation~\eqref{radstresA}, to obtain
\begin{equation}\label{B2apzeta}
    \mathcal{B}_2(a,b) = \frac{b^2a^2\left\{\sigma_a\left[\zeta(1-\Gamma)-2\right]+2\sigma_b^\prime\right\}}{2(b^2-a^2)}.
\end{equation}

As discussed above, we assume that yield first occurs at $r=s=a$ and that the yield function (Equation~\ref{yieldcrit}) is maximised at steady state. These assumptions require that
\begin{equation}\label{yieldA}
    \alpha\sigma^\prime_\theta(a_\mathrm{min})-\sigma^\prime_r(a_\mathrm{min})=y,
\end{equation}
where $a_{\mathrm{min}}$ is the value of $a$ associated with $\sigma_a^\mathrm{min}$. 
Combining Equations~\eqref{genundzeta} and \eqref{yieldA}, we obtain
\begin{equation}\label{yieldB2}
    (\alpha-1)(1-\zeta)\sigma_a+\alpha\left[\frac{2\mathcal{B}_2}{a_\mathrm{min}^2} -\frac{\zeta\sigma_a(1-\Gamma)}{2\ln\left(\frac{b_\mathrm{min}}{a_\mathrm{min}}\right)}\right]=y,
\end{equation}
where $b_\mathrm{min}$ is the value of $b$ associated with $\sigma_a^\mathrm{min}$ and $a_\mathrm{min}$. Evaluating Equation~\eqref{B2apzeta} at $a=a_\mathrm{min}$ and $b=b_\mathrm{min}$ and using the result in Equation~\eqref{yieldB2}, we arrive at
\begin{multline}\label{sigmin}
\sigma_{a}^\mathrm{min}(\zeta; a_\mathrm{min}, b_\mathrm{min}) = \\ \displaystyle\frac{
2\ln\left(\displaystyle\frac{b_\mathrm{min}}{a_\mathrm{min}}\right)\left[y\left(\displaystyle\frac{b_\mathrm{min}^2}{a_\mathrm{min}^2}-1\right)-2\alpha \sigma_b^\prime\left(\displaystyle\frac{b_\mathrm{min}^2}{a_\mathrm{min}^2}\right)\right]
}{
2\displaystyle\ln\left(\frac{b_\mathrm{min}}{a_\mathrm{min}}\right)\left\{ \left(\frac{b_\mathrm{min}^2}{a_\mathrm{min}^2}\right)\Big[\zeta(1-\Gamma\alpha)-\alpha-1\Big]-(\alpha-1)(1-\zeta)\right\} -\alpha\zeta(1-\Gamma)\left( \frac{b_\mathrm{min}^2}{a_\mathrm{min}^2}-1\right)
}.
\end{multline}
For the LL model, we then take $a_\mathrm{min} = a^\mathrm{ref}$ and $b_\mathrm{min} = 1$, at which point $\sigma_{a}^\mathrm{min}$ is fully determined. For the QL model, we take $b_\mathrm{min}=1$ and enforce the kinematic condition at the inner boundary to arrive at an implicit expression for $a_\mathrm{min}$,
\begin{equation}
u_s(a_\mathrm{min})=a_\mathrm{min}-a^\mathrm{ref} = a_\mathrm{min}\left[\frac{(1-\zeta)(1-\alpha\Gamma)\sigma_{a}^\mathrm{min}(a_\mathrm{min}, 1)+y}{\alpha(1-\Gamma^2)}\right].
\end{equation}
For the QQ and NQ models, we enforce kinematic conditions at both boundaries,
\begin{equation}
u_s(a_\mathrm{min})=a_\mathrm{min}-a^\mathrm{ref} = a_\mathrm{min}\left[\frac{(1-\zeta)(1-\alpha\Gamma)\sigma_{a}^\mathrm{min}(a_\mathrm{min}, b_\mathrm{min})+y}{\alpha(1-\Gamma^2)}\right]
\end{equation}
and 
\begin{equation}
u_s(b_\mathrm{min})=b_\mathrm{min}-1 = \frac{b_\mathrm{min}\sigma_b^\prime}{1+\Gamma}+\frac{2\mathcal{B}_2(a_\mathrm{min},b_\mathrm{min})}{b_\mathrm{min}(1-\Gamma^2)}-\frac{\zeta b_\mathrm{min}\sigma_{a}^\mathrm{min}(a_\mathrm{min},b_\mathrm{min})}{2(1+\Gamma)\ln\left(\frac{b_\mathrm{min}}{a_\mathrm{min}}\right)},
\end{equation}
where $\mathcal{B}_2(a_\mathrm{min},b_\mathrm{min})$ and $\sigma_{a}^\mathrm{min}(a_\mathrm{min},b_\mathrm{min})$ are defined in Equations~\eqref{B2apzeta} and \eqref{sigmin}, respectively.

For $\zeta\equiv1$, all of the above can be solved for $a_\mathrm{min}$ explicitly, whereas $b_\mathrm{min}$ must then be determined via numerical root-finding. This is not the case for $\zeta\not\equiv{}1$, in which case the QQ and NQ models require simultaneous root-finding for both $a_\mathrm{min}$ and $b_\mathrm{min}$.

\subsection{Value of $\sigma_a$ at which the material yields completely ($\sigma_a^\mathrm{max}$)}\label{maxsig}

When the elastic-plastic interface reaches the outer radius, $r=s=b$, we enforce the constraint $\sigma_r^\prime(b)=\sigma_b^\prime$ on Equation~(\ref{plasticstressA}) to arrive at
\begin{equation}\label{sigmax}
\sigma_{a}^\mathrm{max}(a_\mathrm{max},b_\mathrm{max})= \frac{\ln\left(\frac{b_\mathrm{max}}{a_\mathrm{max}}\right)\left[\sigma_b^\prime a_\mathrm{max}^K(\alpha-1)-y(a_\mathrm{max}^K-b_\mathrm{max}^K)\right]}{\alpha\zeta(a_\mathrm{max}^K-b_\mathrm{max}^K)+b_\mathrm{max}^K(1-\zeta)(\alpha-1)\ln\left(\frac{b_\mathrm{max}}{a_\mathrm{max}}\right)},
\end{equation}
where $a_\mathrm{max}$ and $b_\mathrm{max}$ are the values of $a$ and $b$, respectively, associated with $\sigma_a^\mathrm{max}$. The values of $a_\mathrm{max}$ and $b_\mathrm{max}$ are then determined by conditions at the inner and outer boundaries. For the LL model, we simply take $a_\mathrm{max}=a^\mathrm{ref}$ and $b_\mathrm{max}=1$ to arrive at an explicit expression for $\sigma_a^\mathrm{max}$. For the QL and QQ models\footnote{This expression also applies to the LL model, but is not needed to determine $\sigma_a^\mathrm{max}$.}, the solution of Equation~\eqref{plast_lin} gives the displacement in the plastic region,
\begin{subequations}\label{up}
\begin{equation}
u_s = \displaystyle \frac{D_1 r}{\beta+1}+\frac{\alpha a D_2}{\beta+\alpha}\left(\frac{r}{a}\right)^\frac{1}{\alpha} + \mathcal{D}_1r^{-\frac{1}{\beta}} \quad a\leq{}r<s,
\end{equation}
where $D_1$ and $D_2$ are as defined in Equation~\eqref{defsA} and continuity of displacement at the elastic-plastic interface leads to an expression for $\mathcal{D}_1$,
\begin{equation}
\mathcal{D}_1 = s^\frac{1}{\beta}\left[u_s(s^+)- \frac{D_1 s}{\beta+1}-\frac{\alpha a D_2}{\beta+\alpha}\left(\frac{s}{a}\right)^\frac{1}{\alpha}\right],
\end{equation}
\end{subequations}
where $u_s(s^+)$ is the displacement from the elastic solution at $r=s$. For the QL model, we take $s=b_\mathrm{max}\equiv1$. We can then calculate $u_s(s=1)$ from the elastic solution (Equation~\ref{genundzeta}), subject to $\sigma_r^\prime(1)=\sigma_b^\prime$ and $\alpha\sigma_\theta^\prime(1) -\sigma_b^\prime=y$. This gives 
\begin{equation}
    u_s(1) =\frac{y+(1-\alpha\Gamma)\sigma_b^\prime}{\alpha(1-\Gamma^2)}, 
\end{equation}
and hence 
\begin{equation}
u_s = \frac{D_1}{\beta+1}\left[r-\left(\frac{1}{r}\right)^\frac{1}{\beta}\right]+\frac{\alpha D_2}{a^K(\beta+\alpha)}\left[r^\frac{1}{\alpha}-\left(\frac{1}{r}\right)^\frac{1}{\beta}\right]+\left[\frac{y+(1-\alpha\Gamma)\sigma_b^\prime}{\alpha(1-\Gamma^2)}\right]\left(\frac{1}{r}\right)^\frac{1}{\beta},
\end{equation}
so that $\sigma_{a}^\mathrm{min}(\zeta)$ is fully determined once $a_\mathrm{max}$ is found via the implicit relation 
\begin{multline}
a_\mathrm{max}-a^\mathrm{ref} = \frac{D_1(a_\mathrm{max})}{\beta+1}\left[a_\mathrm{max}-\left(\frac{1}{a_\mathrm{max}}\right)^\frac{1}{\beta}\right]+\frac{\alpha D_2(a_\mathrm{max})}{a_\mathrm{max}^K(\beta+\alpha)}\left[a_\mathrm{max}^\frac{1}{\alpha}-\left(\frac{1}{a_\mathrm{max}}\right)^\frac{1}{\beta}\right]\\+\left[\frac{y+(1-\alpha)\sigma_b^\prime}{\alpha(1-\Gamma^2)}\right]\left(\frac{1}{a_\mathrm{max}}\right)^\frac{1}{\beta}.
\end{multline}

For the QQ model, rigorous treatment of the kinematics gives $u_s(b) = b-1$ and hence
\begin{multline}
u_s = \frac{D_1}{\beta+1}\left[r-b_\mathrm{max}\left(\frac{b_\mathrm{max}}{r}\right)^\frac{1}{\beta}\right]+\frac{\alpha  D_2}{a^K(\beta+\alpha)}\left[r^\frac{1}{\alpha}-b_\mathrm{max}^\frac{1}{\alpha}\left(\frac{b_\mathrm{max}}{r}\right)^\frac{1}{\beta}\right]\\ +(b_\mathrm{max}-1)\left(\frac{b_\mathrm{max}}{r}\right)^\frac{1}{\beta}.
\end{multline}
To find $b_\mathrm{max}$, we once again appeal to the elastic problem at $r=s=b$. Solving Equation~\eqref{genundzeta} subject to $\sigma_r^\prime(b_\mathrm{max})=\sigma_b^\prime$ and $ \alpha\sigma_\theta^\prime(b_\mathrm{max})-\sigma_b^\prime=y$ and equating the result with $u_s(b_\mathrm{max}) = b_\mathrm{max}-1$, we arrive at
\begin{subequations}
    \begin{equation}\label{bmaxQQapzeta}
        b_\mathrm{max} = \frac{\alpha(1-\Gamma^2)}{\alpha(1-\Gamma^2)+(\alpha\Gamma-1)\sigma_b^\prime-y}.
\end{equation}
Note that this result also applies to the NQ model, as we have not used the plastic flow law in its calculation. Hence, $\sigma_{a}^\mathrm{min}(\zeta)$ is fully determined once $a_\mathrm{max}$ is found via the implicit relation 
\begin{multline}
a_\mathrm{max}-a^\mathrm{ref}= \frac{D_1(a_\mathrm{max}, b_\mathrm{max})}{\beta+1}\left[a_\mathrm{max}-b_\mathrm{max}\left(\frac{b_\mathrm{max}}{a_\mathrm{max}}\right)^\frac{1}{\beta}\right]+\\\frac{\alpha D_2(a_\mathrm{max}, b_\mathrm{max})}{a_\mathrm{max}^K(\beta+\alpha)}\left[a_\mathrm{max}^\frac{1}{\alpha}-b_\mathrm{max}^\frac{1}{\alpha}\left(\frac{b_\mathrm{max}}{a_\mathrm{max}}\right)^\frac{1}{\beta}\right]+(b-1)\left(\frac{b_\mathrm{max}}{a_\mathrm{max}}\right)^\frac{1}{\beta},
\end{multline}
\end{subequations}
with $b_\mathrm{max}$ a known constant as defined in Equation (\ref{bmaxQQapzeta}). For the NQ model, Equation~\eqref{govern21A} cannot be solved analytically, prohibiting the derivation of an algebraic expression for $a_\mathrm{max}$.

\subsection{Implicit analytical solution to the LL model}
\label{LLk0}

For the LL model, we derive analytical expressions for the displacement in both the elastic region ($s<r<b\equiv1$) and the plastic region ($a\equiv a^\mathrm{ref}<r<s$), coupled with an implicit expression for the value of $s$. Solving the problem defined by Equations~\eqref{plast_lin} and \eqref{elasticA} subject to boundary conditions \eqref{cont_u}--\eqref{kin_b} leads to
\begin{subequations}\label{LL_soln}
\begin{equation}
    u_s= \left\{ \begin{array}{c}
\displaystyle \displaystyle \frac{D_1 r}{\beta+1}+\frac{\alpha a D_2}{\beta+\alpha}\left(\frac{r}{a^\mathrm{ref}}\right)^\frac{1}{\alpha} + \mathcal{D}_1r^{-\frac{1}{\beta}}, \quad a^\mathrm{ref}<r<s \\
\\ 
\displaystyle -\frac{\sigma_a\zeta r\ln(r) }{2\ln\left({a^\mathrm{ref}}\right)}+\frac{\mathfrak{D}_1r}{1+\Gamma}+\frac{\mathfrak{D}_2}{(1-\Gamma)r} + \frac{\zeta\sigma_ar}{2(1+\Gamma)\ln\left(a^\mathrm{ref}\right)}, \quad s\leq r<1 \\
\end{array}\right.,
\end{equation}
where $D_1(s)$ and $D_2(s)$ are as defined in Equation~\eqref{defsA},
\begin{equation}
\mathcal{D}_1(s)\equiv s^\frac{1}{\beta}\left[\underbrace{-\frac{\sigma_a\zeta s\ln(s) }{2\ln\left({a^\mathrm{ref}}\right)}+\frac{\mathfrak{D}_1s}{1+\Gamma}+\frac{\mathfrak{D}_2}{(1-\Gamma)s} +\frac{\zeta\sigma_as}{2(1+\Gamma)\ln\left({a^\mathrm{ref}}\right)}}_{u_s(s^+)}- \frac{D_1 s}{\beta+1}-\frac{\alpha a D_2}{\beta+\alpha}\left(\frac{s}{a}\right)^\frac{1}{\alpha}\right],
\end{equation}
\begin{equation}
\mathfrak{D}_1(s)\defeq \mathfrak{D}_2 +\sigma_b^\prime,
\end{equation}
and
\begin{equation}
\mathfrak{D}_2(s)\defeq \frac{s^2\left\{2\ln(a^\mathrm{ref})[y-\sigma_b^\prime(\alpha-1)]-\alpha\zeta\sigma_a(1-\Gamma)+\zeta\sigma_a(\alpha-1)(1+\Gamma)\ln(s)\right\}}{2\ln(a^\mathrm{ref})[1+\alpha+(\alpha-1)s^2]}.
\end{equation}
\end{subequations}
The value of $s$ is then determined via the implicit expression
\begin{equation}\label{solve_s_LL}
C_1+C_2\left(\frac{s}{a^\mathrm{ref}}\right)^K-\sigma_b^\prime+\frac{(1+\Gamma)\sigma_a\zeta\ln(s)}{2\ln(a^\mathrm{ref})}+\frac{1-s^2}{s^2}\mathcal{B}_2= 0,
\end{equation}
where $C_1(s)$ and $C_2(s)$ are as defined in Equation~\eqref{defsAA} and $\mathcal{B}_2$ is as defined in Equation~\eqref{B2apzeta}. Note that $s$ can take any value from $a^\mathrm{ref}$ to 1. We solve Equation~\eqref{solve_s_LL} for $s$ via numerical root-finding using \verb+MATLAB+'s \verb+fzero+. The solution is then fully prescribed by Equation~\eqref{LL_soln}.

\subsection{Numerical method for the QL, QQ and NQ models}
\label{numerics}

For all models, we have a closed, coupled, free-boundary BVP in terms of $u_s$, as presented in Equations~\eqref{govern21A} for the NQ model and Equations~\eqref{plast_lin} and \eqref{elasticA}--\eqref{kin_b} for the other three models. For all models, the governing ODE in the elastic region (Equation~\ref{elasticA}) can be solved analytically. For the LL, QL, and QQ models, the governing ODE in the plastic region (Equation~\ref{plast_lin}) can also be solved analytically. However, doing so leads to a strongly nonlinear root-finding problem for $\{a,s,b\}$, in which we do not have a good initial guess for $s$. We avoid this by instead solving all of these models and the NQ model numerically by adapting the Chebyshev spectral collocation method of \citet{auton2017arteries, auton2018arteries}.

To do so, we map the plastic region ($r\in[a,s]$) and the elastic region ($r\in[s,b]$) to separate Chebyshev grids, each with $N$ Chebyshev nodes. Note that both domains include the elastic-plastic interface ($r=s$), meaning that the full domain $r\in[a,b]$ is discretised by $2N-1$ nodes (``collocation points''). We then discretise all derivatives using dense Chebyshev differentiation matrices, thus converting this coupled free-boundary BVP into a system of algebraic equations. We solve for $2N$ unknowns, comprising the displacement at the $2N-1$ nodes as well as the value of $s$; we therefore require a system of $2N$ constraints.

In the plastic region, the governing ODE (Equation~\ref{plasticA}) is first order and thus must be enforced at exactly $N-1$ nodes. We enforce this for the first $N-1$ nodes of the discretised domain ($r\in[a,s)$), having already used the boundary condition $\sigma_r^\prime(a)=(1-\zeta)\sigma_a$ in the derivation this ODE. In the elastic region, the governing ODE (Equation~\ref{elasticA}) is second order and must be enforced at exactly $N-2$ nodes. We enforce this for the $N-2$ interior nodes ($r\in(s,b)$) and then impose $\sigma_r^\prime(b)=\sigma_b$ at the outer boundary (Equation~\ref{outer_BC}). Lastly, we enforce two conditions at $r=s$: Continuity of radial effective stress (Equation~\ref{another_label}) and the requirement that the elastic stresses must satisfy the yield condition (Equation~\ref{yield_in_elastic}). We solve this nonlinear system via Newton iteration. At each iteration, we update the domain via the kinematic conditions at the inner and outer boundaries (Equations~\ref{kin_a} and \ref{kin_b}) and the current value of $s$.

The main structural difference between the models presented here and the models presented in \citet{auton2017arteries, auton2018arteries} is the coupling of an elastic domain with a plastic domain at a free (but non-material) boundary. As a result, the derivation of an exact analytical Jacobian matrix is nontrivial and we instead approximate the Jacobian matrix numerically. In addition, whereas $a^\mathrm{ref}$ and $b^\mathrm{ref}\equiv{}1$ can be used as reasonable initial guesses for $a$ and $b$, no such guess exists for $s$. To accommodate the approximate nature of the Jacobian and the lack of a good initial guess for $s$, we begin by calculating the solution for the LL model for a given value of $\zeta$ and a given, small value of $\sigma_a<\sigma_a^\mathrm{min}$. We use this solution as an initial guess for the numerical solution of the QL model, which then provides an initial guess for the QQ model, which then provides an initial guess for the NQ model. We extend the solution for each model to larger values of $\sigma_a>\sigma_a^\mathrm{max}$ using numerical continuation. Note that $\sigma_a^
\mathrm{max}<\sigma_a^\mathrm{min}<0$. This approach allows for the fact that the model behaviours diverge from one another as the driving strength increases (see Figure~\ref{fig6-2}). The same approach can also be used for increasing $\zeta$ at fixed $\sigma_a$, since deformation increases with $\zeta$ (see Figures~\ref{fig6-3} and \ref{fig6-4}).

\section{Results}
\label{results}

We now use our models and solutions to study the poroelasto-plastic deformation of a cohesive granular cylinder, subject to pressurisation of the inner cavity. As discussed in \S\ref{Eqn_sum_NQL} and \S\ref{s:solutions}, we have developed solutions for four distinct model combinations: LL, QL, QQ and NQ. We fix all parameters based on the values discussed in \S\ref{params} below, except for $\sigma_a$ (or $q$) and $\zeta$, which we vary.

For comparison with our poroelasto-plastic models, we additionally extend the purely poroelastic models from \citet{auton2017arteries, auton2018arteries} by modifying the boundary conditions to incorporate the permeability-load parameter $\zeta$. Specifically, we present solutions to the L-$k_0$ model (here denoted by `L'; linear elasticity, linearised kinematics, and constant permeability) and the Q-$k_0$ model (here denoted by `Q'; linear elasticity, rigorous kinematics, and constant permeability) alongside our poroelasto-plastic solutions in Figures~\ref{fig6-1}--\ref{fig6-4} below. This comparison illustrates the impact of plasticity across different values of $\zeta$.

To investigate the impact of fluid injection into the pre-stressed subsurface, we wish to isolate the additional stress and deformation due to fluid injection from those due to background compression. We approximate the disturbance in all quantities due to fluid injection by subtracting their values in the compressed initial state (III in Figure~\ref{fig1}) from their values in the final steady state (V in Figure~\ref{fig1}). For the displacement, for example, we consider the disturbance $\delta{u_s}\defeq u_s -u_{s,0}$. Recall that these compressed initial states are presented in \S\ref{s:ref_to_0}. We subtract the L initial state from the LL and QL models, and the Q initial state from the QQ and NQ models.

\subsection{Parameters}\label{params}

We adopt a set of parameter values motivated by boreholes in the subsurface. As such, we adopt $\breve{\mathcal{M}}\sim 50$~GPa and $\breve{\Lambda}\sim 27$~GPa (Young modulus $\breve{E}\sim25$~GPa and Poisson ratio $\nu\sim 0.36$) and a porosity of $\phi_f^\mathrm{ref}\sim 0.20$ as typical properties of sedimentary rocks such as sandstones and shales~\citep[\textit{e.g.},][]{goodman1989introduction, hart1995laboratory, bobko2008nano, bobko2011nanogranular, rickman2008practical, britt2009geomechanics}. We further assume moderate internal friction and cohesion, $\varphi \sim 35\text{\textdegree}$ and $\breve{c}\sim 120$~MPa, but very little dilation, $\psi \sim 0.3\text{\textdegree}$ \cite{mandl2005rock, bobko2008nano, bobko2011nanogranular, vermeer1984non}. We take $-\breve{\sigma}_b^\prime\sim 50$~MPa, as appropriate for a depth of $\sim$$2.5\,\mathrm{km}$~\citep{neuzil1994permeable, islam2010stability}. These dimensional values correspond to $\Gamma =0.55$, $\alpha=4$, $\beta =1.01$, $y=10^{-2}$, $a^\mathrm{ref} =10^{-4}$, $\phi_f^\mathrm{ref} = 0.2$ and $\sigma_b^\prime=-1\times 10^{-3}$. Note that $\alpha\nu \equiv \frac{\alpha\Gamma}{1+\Gamma}\sim 1.4>1$ and $y(1-a^2)+2\sigma_b^\prime \sim 0.008>0$, where we take $a\approx{}a^\mathrm{ref}$; as such, the relevant yield condition is $\mathcal{F}_{\theta,r}=0$ (see Appendix~\ref{which_yield}). Below, we consider the response of this material to different driving strengths (varying $\sigma_a$ or $q$), and we additionally consider the transition from impermeable to fully permeable by varying $\zeta$ from 0 to 1.

\subsection{Fully permeable: fixed $q$}\label{s:figs_q}

\begin{figure*}
    \vspace{-1cm}
    \begin{center}
        \includegraphics[width=1.0\textwidth]{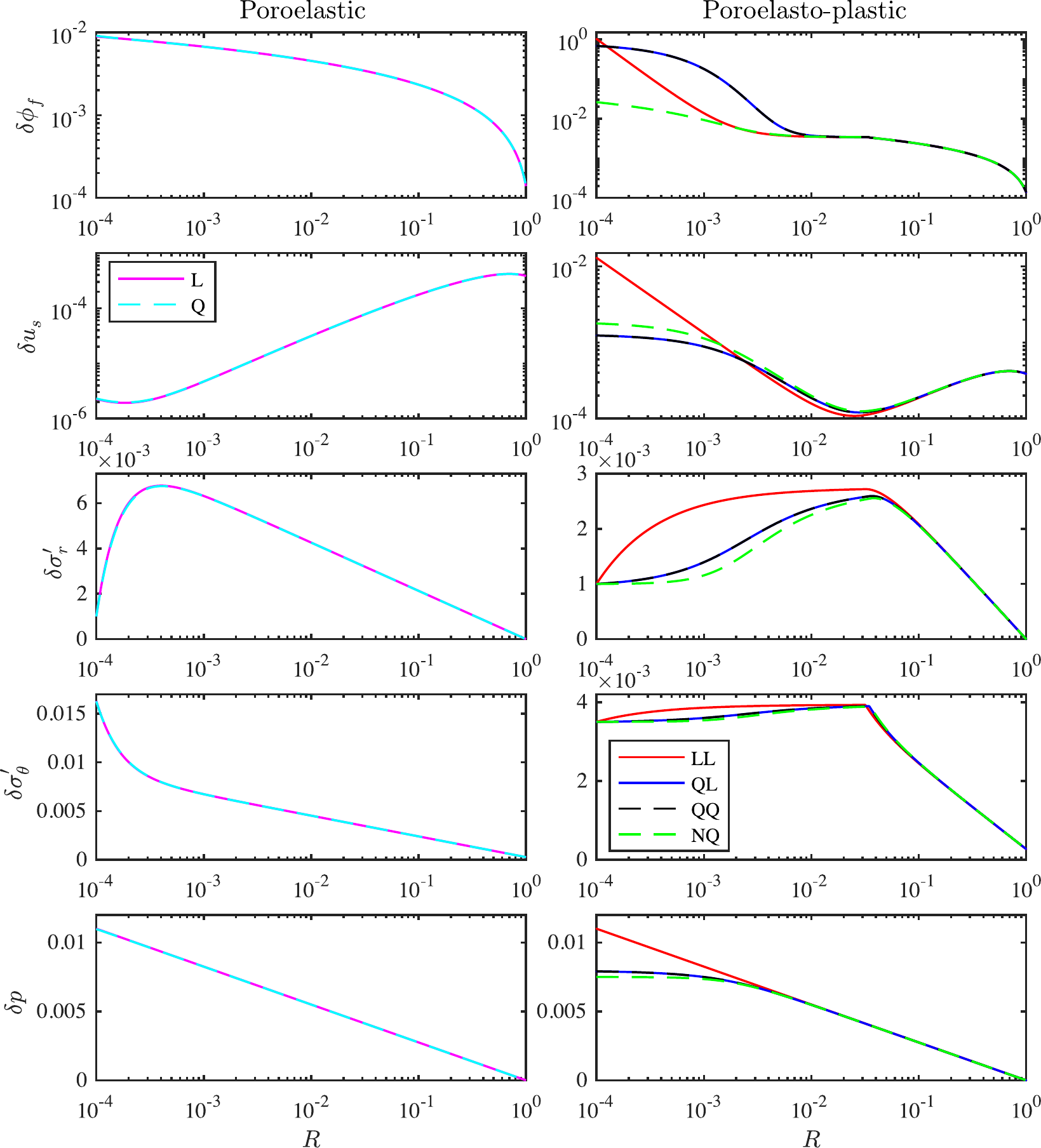}
    \end{center}
    \vspace{-1cm}
    \caption{\small{Two poroelastic models (left column) and four poroelasto-plastic models (right column), all fully permeable ($\zeta\equiv{}1$), at steady state for fixed $q\approx0.0012$ ($s\approx{}0.03$). We show the disturbances due to fluid injection relative to a compressed initial state (III$\mapsto$V in Fig.~\ref{fig1}): $\delta{\phi}$ (first row), $\delta{u_s}$ (second row), $\delta{\sigma_r^\prime}$ (third row), $\delta{\sigma_\theta^\prime}$ (fourth row), and $\delta{p}$ (last row). For clarity, we plot these results against the Lagrangian coordinate $R(r) = r -u_s$ and on a logarithmic horizontal scale. Note that the top two rows are also on a logarithmic vertical scale. Plasticity enables large deformations in the plastic region that amplify the importance of model choice there; model choice is unimportant in the purely poroelastic case, and in the elastic region of the poroelasto-plastic case. For reference, we plot the true steady-state values of the same quantities (\textit{i.e.}, without subtracting the initial state) in Figure~\ref{figA-noref} in Appendix~\ref{Figs}.} \label{fig6-1} }
\end{figure*}

\begin{figure*}
    \begin{center}
        \includegraphics[width=1.0\textwidth]{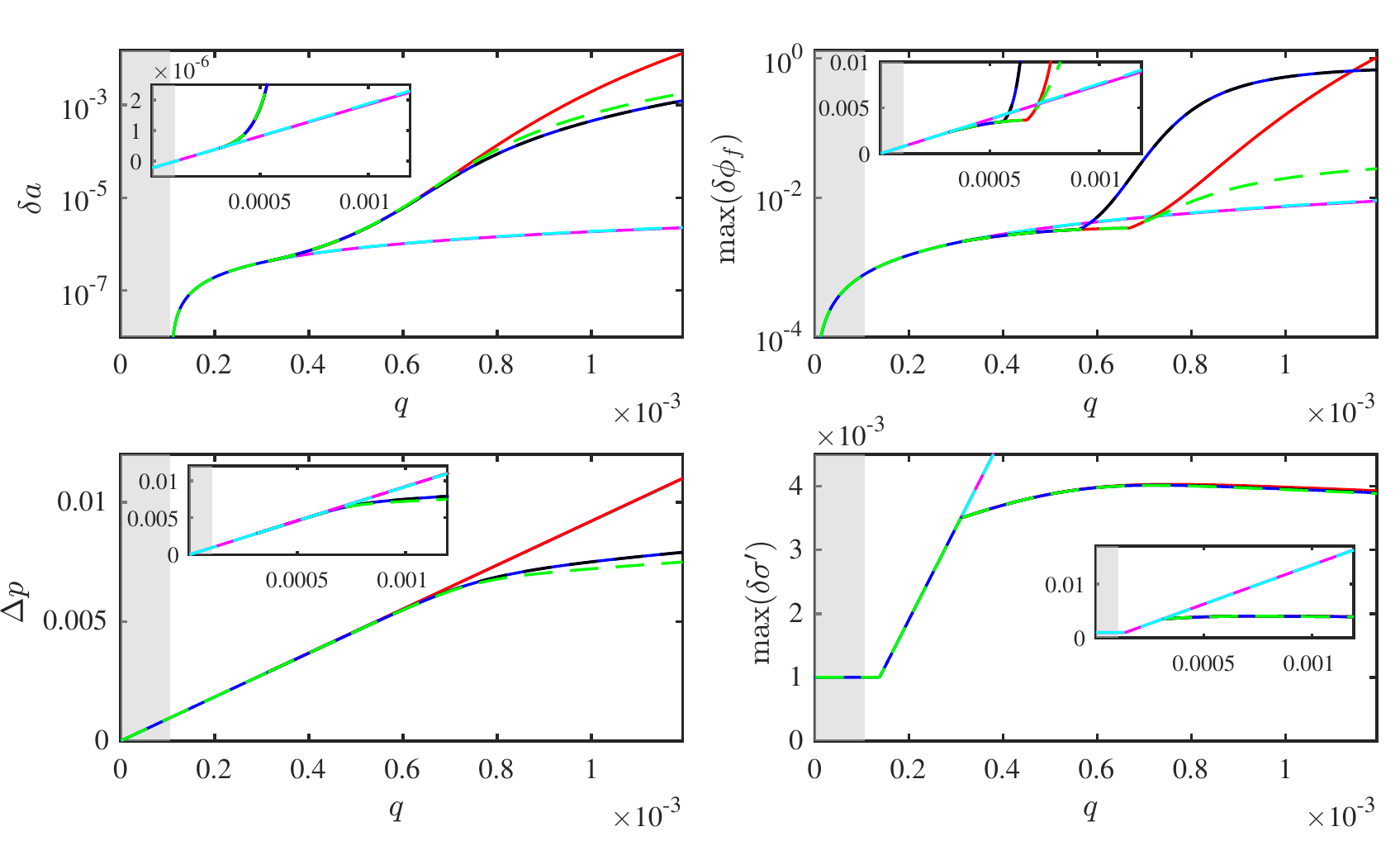}
    \end{center}
    \caption{\small{Four summary quantities against $q$ for the fully permeable case ($\zeta\equiv1$), with the same colours as in Figure~\ref{fig6-1}. We plot the disturbance to the inner radius $\delta{a}$ (top left), the maximum disturbance in porosity $\mathrm{max}(\delta{\phi_f})$ (top right), the injection pressure $\Delta{p}$ (bottom left), and the maximum disturbance in effective stress $\mathrm{max}(\delta{\sigma^\prime})$ (bottom right). We focus on the poroelasto-plastic models in the main plots and on the poroelastic models in the insets. The shaded grey regions indicate the values of $q$ for which the inner cavity contracts. These results again highlight that plasticity leads to a large deviation from poroelastic behaviour, and that model choice is unimportant for the poroelastic models but very important for the poroelasto-plastic models---particularly with regard to the kinematics in the plastic region. Note that yield first occurs for $q\approx{}3.1\times{}10^{-4}$ (the second corner in $\mathrm{max}(\delta{\sigma^\prime})$).} \label{fig6-2} }
\end{figure*}

We begin comparing poroelasto-plastic behaviour with purely poroelastic behaviour in the context of a fully permeable material ($\zeta\equiv{}1$; Figures~\ref{fig6-1} and \ref{fig6-2}). Recall that, for this model, we drive the system with a fixed flow rate $q$ instead of a fixed pressure difference $\Delta{p}\equiv-\sigma_a$ (see \S\ref{s:inner_bc}). For each model, a given value of $q$ will correspond to a specific steady-state value of $\Delta{p}$; however, this value will be different for each model. Note that, as with $\sigma_a$, there exists a minimum value $q_\mathrm{min}$ at which yield first occurs and a maximum value $q_\mathrm{max}$ at which the material yields completely. These values can be calculated via $q_\mathrm{min} = -\sigma_{a}^\mathrm{min}(1)/\ln\left(\frac{b}{a}\right)$ and $q_\mathrm{max} = -\sigma_{a}^\mathrm{max}(1)/\ln\left(\frac{b}{a}\right)$, where $\sigma_{a}^\mathrm{min}(1)$ and $\sigma_{a}^\mathrm{max}(1)$
are defined in Equations~\eqref{sigmin} and \eqref{sigmax}, respectively.

In Figure~\ref{fig6-1}, we consider $q = 0.0012$. For these parameters, the LL model has $q_\mathrm{min}\approx 3.1\times10^{-4}$ and $q_\mathrm{max} \approx 0.0033$, so that our chosen value of $q$ is about $4q_\mathrm{min}$ and about $q_\mathrm{max}/3$. For this value of $q$, we plot the disturbance due to fluid injection for various key quantities for the poroelastic models (left column) and the poroelasto-plastic models (right column) against the Lagrangian radial coordinate $R(r)\equiv{}r-u_s(r)$. For reference, we plot the true steady-state values of the same quantities (\textit{i.e.}, without subtracting the initial state) in Figure~\ref{figA-noref} in Appendix~\ref{Figs}. Note that $\delta{p}\equiv{}p$ because $p_0\equiv{}0$ for all models.

Although this value of $q$ is substantially higher than $q_\mathrm{min}$, it leads to relatively small elastic deformations in the absence of yield. As a result, the predictions of the two purely poroelastic models are essentially indistinguishable (Figure~\ref{fig6-1}, left column). The disturbances $\delta{\phi_f}$, $\delta{\sigma_\theta^\prime}$, and $\delta{p}$ all have maxima at the cavity wall, with $\delta{\sigma_\theta^\prime}$ falling off relatively steeply from this value as $R$ increases. In contrast, $\delta{u_s}$ and $\delta{\sigma_r^\prime}$ have internal maxima near the outer boundary ($R\sim0.7$) and near the inner boundary ($R\sim{}4\times{}10^{-4}$), respectively. The value of $\delta{\sigma_r^\prime}$ is pinned to $-\sigma_b^\prime$ at the inner boundary by construction since $\sigma_r^{\prime}(a)=0$ and $\sigma_{r,0}^\prime\equiv\sigma_b^\prime$, and vanishes at the outer boundary since $\sigma_r^{\prime}(b)=\sigma_{r,0}^\prime=\sigma_b^\prime$. Note that $\delta{\phi_f}$ is strictly positive, meaning that the porosity everywhere has increased relative to the compressed initial state.

Upon introducing plastic yield (Figure \ref{fig6-1}, right column), all quantities except $\delta{p}$ behave drastically differently than in the purely poroelastic case. Most importantly, plasticity enables much larger displacements in the plastic region (near the cavity wall), with much smaller associated stresses. The strains in the plastic region are no longer infinitesimal---for example, $\mathrm{max}\left(\frac{u_s}{r}\right)\approx 130$ for the poroelasto-plastic LL model whereas $\mathrm{max}\left(\frac{u_s}{r}\right)\approx 0.02$ for the poroelastic L and Q models, with this maximum occurring at the inner boundary in both cases. These large strains make model choice much more important in the plastic region, as evidenced by the substantial differences between the models there. In the elastic region, however, the maximum strain remains relatively small---for example, $\mathrm{max}\left(\frac{u_s^\mathrm{e}}{r}\right)\approx 0.003$ for the poroelasto-plastic LL model. This suggests that it is reasonable to linearise the elastic part of the solution, even in the presence of large displacements in the plastic region. This idea is further supported by the other results in Figures~\ref{fig6-1}--\ref{fig6-4}, in which the QL and QQ models are essentially indistinguishable. We conclude that model choice in the elastic region is relatively unimportant to the overall behaviour. Note also that although model choice in the plastic region is important for the behaviour in the plastic region, the resulting (substantial) differences in behaviour have relatively little impact on behaviour in the elastic region.

In the plastic region, the LL model generally predicts the most extreme behaviour, with additional facets of nonlinearity increasingly moderating this behaviour. The QL and QQ models are effectively indistinguishable from each other, again demonstrating that model choice in the elastic region is unimportant. The NQ model predicts very similar behaviour to the QL and QQ models except with regard to porosity, where the NQ predicts much smaller values of $\delta{\phi_f}$ than any of the models. This discrepancy is due to the fact that $\mathrm{d}u_s/\mathrm{d}r$ is much more negative near the inner radius for the NQ model than for the QQ or QL models (Eq.~\ref{eq:phi_to_u}). The LL model predicts $\delta{\phi_f}>1$ near the inner boundary, highlighting the fact that linearising the kinematics in the plastic region can lead to nonphysical solutions\footnote{In fact, we need $\delta{\phi_f}\lesssim 0.8$ since $\delta{\phi_f}\equiv \phi_f-\phi_{f,0}\approx{}0.8$ for these parameters (\textit{cf.} Figure \ref{figA-noref}).} (see Figure \ref{figA-unreal} in Appendix~\ref{Figs}). Note that, despite the linearisation of the kinematics in the elastic region, the QL model does not have this feature. 

The internal maximum in $\delta{\sigma_r^\prime}$ has shifted further into the material and decreased in magnitude relative to the poroelastic case, now occurring at the elastic-plastic interface and with a magnitude of about $1/2$ of the poroelastic value. The maximum value of $\delta{\sigma_\theta^\prime}$ is now internal rather than at the inner radius, also occurring at the elastic-plastic interface and with a magnitude of about $1/4$ of the poroelastic value. This quantity has a sharp transition across the elastic-plastic interface, meaning that $\frac{\mathrm{d}}{\mathrm{d}r}(\delta{\sigma_\theta^\prime})$ is discontinuous across $r=s$. It is straightforward to show from the boundary conditions at $r=s$, the yield condition, and mechanical equilibrium that $\delta{u_s}$, $\delta{\sigma_r^\prime}$, $\frac{\mathrm{d}}{\mathrm{d}r}(\delta{\sigma_r^\prime})$, $\delta{\sigma_\theta^\prime}$, and $\delta{p}$, and $\frac{\mathrm{d}}{\mathrm{d}r}(\delta{p})$ must all be continuous across $r=s$. However, $\frac{\mathrm{d}}{\mathrm{d}r}(\delta{u_s})$ and therefore also $\delta{\phi_f}$ are weakly discontinuous across $r=s$ (not readily visible in the figure). The discontinuity in $\frac{\mathrm{d}}{\mathrm{d}r}(\delta{\sigma_\theta^\prime})$ is particularly prominent because it occurs at the maximum value of $\sigma_\theta^\prime$, whereas those in
$\delta{\phi_f}$ and $\frac{\mathrm{d}}{\mathrm{d}r}(\delta{u_s})$ occur in regions where $\delta{\phi_f}$ and $\delta{u_s}$ are themselves small.

For the poroelastic L and poroelasto-plastic LL models, $\delta{p}$ decreases exactly logarithmically with $r$ because $\frac{\mathrm{d}p}{\mathrm{d}r}=-\frac{q}{r}$, and these linearised models do not account for the moving boundaries. The poroelastic Q model exhibits essentially the same behaviour since the poroelastic displacements are small. The poroelasto-plastic QL, QQ, and NQ models do account for the substantial increase in $a$ and the comparatively small increase in $b$ due to fluid injection, leading to lower injection pressures at steady state.

In Figure~\ref{fig6-2}, we present results for a wide range of $q$ in terms of four summary quantities. Note that, for values of $q$ for which yield does not occur (\textit{i.e.}, $q<q_\mathrm{min}$), the LL and QL models behave according to the L model, while the QQ and NQ models behave according to the Q model. The maximum value of $q$ shown in Figure~\ref{fig6-2} is $q\approx 0.0012$, which is the value used in Figure~\ref{fig6-1}.

For $q\lesssim{}10^{-4}$ (Figure~\ref{fig6-2}, grey band), the disturbance in the inner radius $\delta{a}$ is negative (Figure~\ref{fig6-2}, top left). This contraction occurs because $q$ is not large enough to support the compressive confining stress, such that the hypothetical casing would need to partially support the material to enforce $a\geq{}a_0$ (II to III in Fig.~\ref{fig1}). Since we do not model the casing, our results in this range produce a steady state that is inconsistent with state III in Fig.~\ref{fig1}. For larger values of $q$, the inner radius increases monotonically with $q$ as injection increasingly pushes the material radially outwards. After yield occurs, the poroelasto-plastic models deform increasingly more than the poroelastic models, and the ordering of the models is the same as in Figure~\ref{fig6-1}.

For small flow rates, the maximum porosity disturbance $\mathrm{max}(\delta{\phi_f})$ is small (Figure \ref{fig6-2}, top right). Much like $\delta{a}$, $\mathrm{max}(\delta{\phi_f})$ increases linearly with $q$ until yield and then departs strongly from linear behaviour for the poroelasto-plastic models. For a small range of $q$ after yield, the poroelasto-plastic models predict values of $\mathrm{max}(\delta{\phi_f})$ that are less than the corresponding poroelastic predictions, which occurs because yield reduces the maximum tensile stress and relatively little plastic flow has occurred. For larger values of $q$, the predictions of the poroelasto-plastic models are much larger than those of the poroelastic models as deformations grow larger and plastic flow leads to significant dilation. The NQ model predicts a much weaker departure from poroelasticity than the other three poroelasto-plastic models.

The injection pressure $\Delta p\equiv{}p(a)\equiv\delta{p(a)}$ increases linearly with $q$ before yield, and continues to increase linearly with $q$ for the LL model after yield (Figure \ref{fig6-2}, bottom left). For the other poroelasto-plastic models, $\Delta{p}$ increases somewhat slower than linearly as the inner radius increases. However, this is a weak effect because the changes in inner and outer radii are ultimately small for all $q$. The injection pressure is important because it is one of the few observables during subsurface operations (\textit{i.e.}, it can be measured from the surface in realtime).

For small $q$, the maximum effective stress disturbance $\max(\delta{\sigma^\prime})$ is constant and equal to $-\sigma_b^\prime=10^{-3}$ (Figure \ref{fig6-2}, bottom right). This is an artefact of the fact that $\sigma_a^\prime\equiv0$, such that $\delta{\sigma_a^\prime}\equiv\sigma_a^\prime-\sigma_b^\prime=10^{-3}$. This is the maximum stress (and the maximum stress disturbance) for small $q$ because all of the other stresses remain compressive until $q$ becomes large enough to generate tensile stresses. Note again that these values are only physically meaningful for $q\gtrsim{}10^{-4}$ (outside the grey region). All models exhibit a corner in $\max(\delta{\sigma^\prime})$ before yield due to a change in which stress component exhibits this maximum: $\max(\delta{\sigma^\prime})$ is equal to $\delta{\sigma_r^\prime}(a)$ for $q\lesssim{}1.4\times{}10^{-4}$ and to $\delta{\sigma_\theta^\prime}(a)$ for $q\gtrsim{}1.4\times{}10^{-4}$. The poroelasto-plastic models exhibit a second corner at yield ($q\approx{}3.1\times{}10^{-4}$) due to a change in the location of this maximum: $\max(\delta{\sigma^\prime})$ is equal to $\delta{\sigma_\theta^\prime}(a)$ before yield and to $\delta{\sigma_\theta^\prime}(s)$ after yield.

In Figure~\ref{figA-sig} in Appendix~\ref{Figs}, we plot the same four quantities as in Figure~\ref{fig6-2}, but against $-\sigma_b^\prime$ for fixed $q$. This shows the transition from an unconstrained cylinder (no confining stress at the outer boundary) to a strongly compressed cylinder, providing a link with results of \citet{auton2017arteries, auton2018arteries} for an unconstrained, fully permeable poroelastic cylinder. Within each class of models (poroelastic and poroelasto-plastic), the ordering of the models is preserved for each quantity as $-\sigma_b^\prime$ varies. Additionally, all quantities converge smoothly towards poroelastic behaviour as $-\sigma_b^\prime$ increases. As a result, it seems reasonable to extrapolate the conclusions of \cite{auton2017arteries, auton2018arteries} for a poroelastic cylinder with deformation-dependent permeability to the poroelasto-plastic cylinders considered here. Figure~2a of \citet{auton2017arteries} suggests that deformation-dependent permeability leads to a much smaller $\Delta{p}$ for a given value of $q$ at steady state, and thus to less deformation and lower stresses. We expect that the same would be true for the poroelasto-plastic scenarios considered here.

\subsection{Impermeable to fully permeable: fixed $\sigma_a$}
\label{s:figs_zeta}

\begin{figure*}[tbp]
    \begin{center}
        \includegraphics[width=1.0\textwidth]{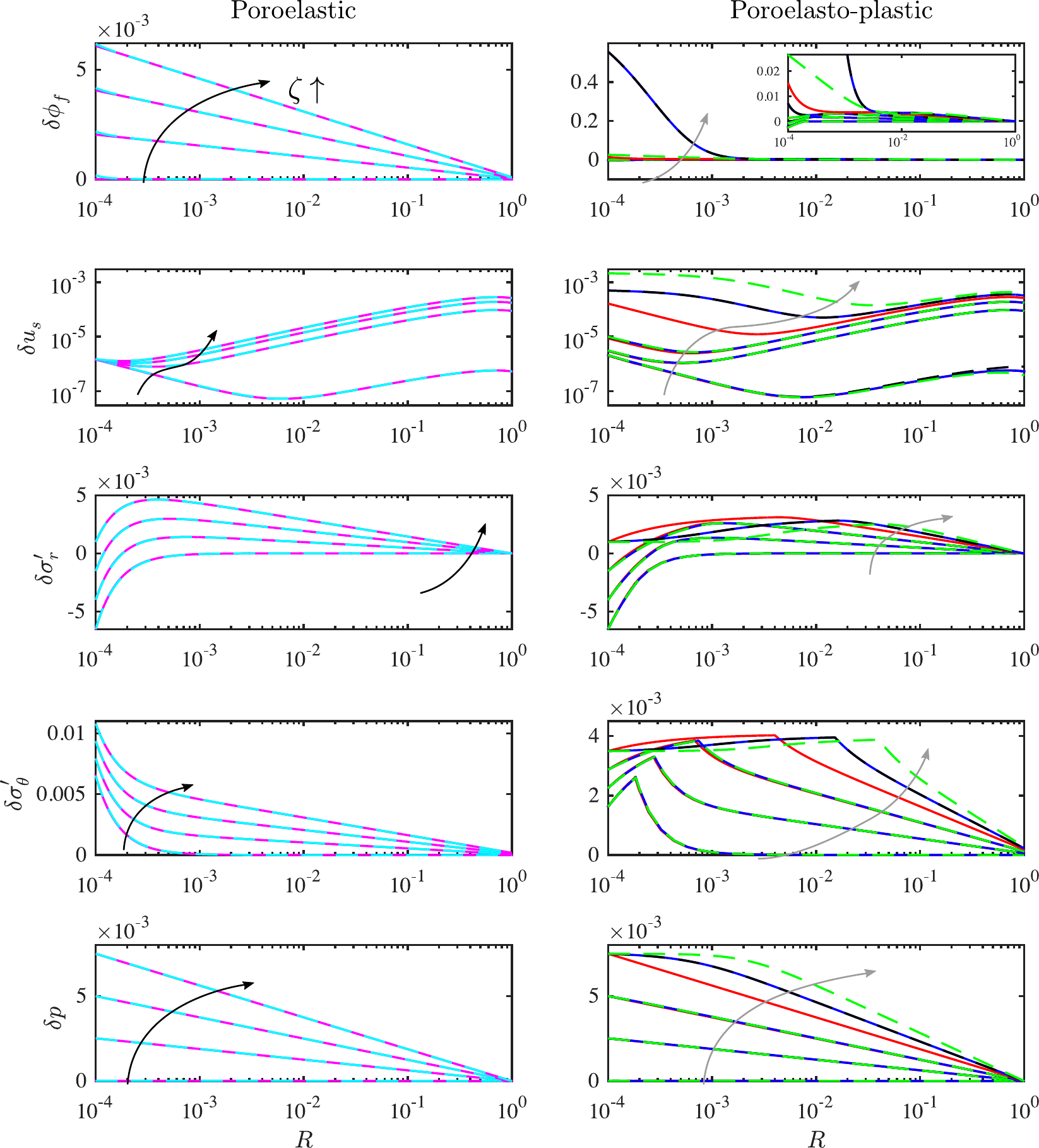}
    \end{center}
    \vspace{-0.4cm}
    \caption{\small{As in Figure \ref{fig6-1}, we plot two poroelastic models (left column) and four poroelasto-plastic models (right column) at steady state, here for fixed $\sigma_a=-0.0075$ and $\zeta\in\{0.002, 0.335, 0.665,0.998\}$. Colours are the same as in Figures~\ref{fig6-1} and \ref{fig6-2}. For the poroelastic models, all quantities increase with $\zeta$ and there is again no discernible difference between the models. For the poroelasto-plastic models, all quantities increase with $\zeta$ to a point, after which the effective stresses from the more nonlinear models become weakly nonmonotonic in $\zeta$. The difference between models grows with $\zeta$ as the deformations grow larger (except for the QL and QQ models, which are again indistinguishable).} \label{fig6-3} }
\end{figure*}

\begin{figure*}
    \begin{center}
    \includegraphics[width=1.0\textwidth]{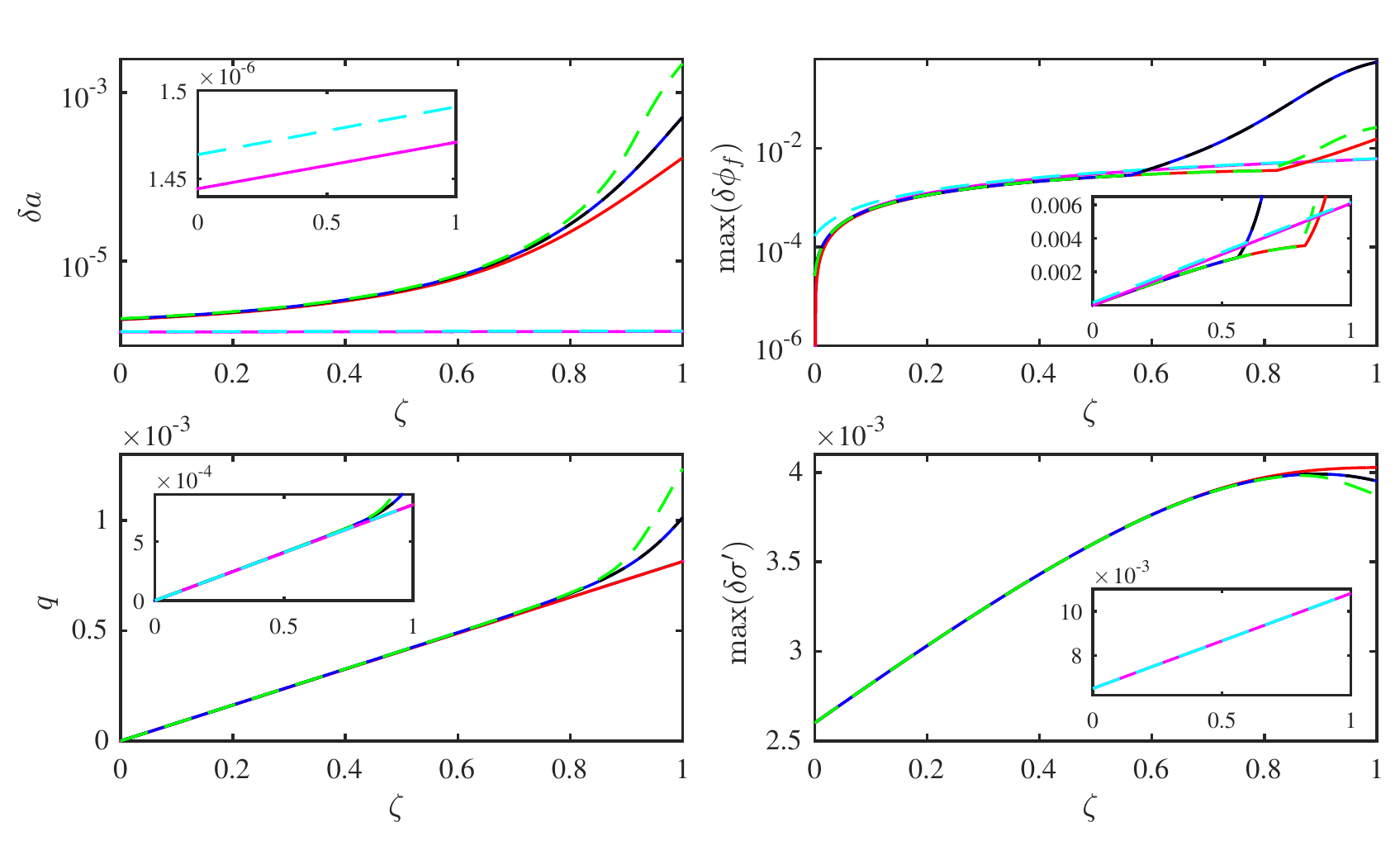}
    \end{center}
    \caption{\small{As in Figure~\ref{fig6-2}, we plot four summary quantities, now against $\zeta$ for fixed $\sigma_a = -0.0075$. We again focus on the poroelasto-plastic models in the main plots and on the poroelastic models in the insets. Colours are the same as in Figures~\ref{fig6-1}--\ref{fig6-3}. Note that plasticity enables much larger displacements, increasingly so as $\zeta$ increases. Whereas $\delta{a}$, $\max(\delta{\phi_f})$, and $q$ increase monotonically with $\zeta$ for all models, $\max(\delta{\sigma^\prime})$ is nonmonotonic in $\zeta$ with an internal maximum near $\zeta=1$ for the QL, QQ, and NQ models. Note that, for $q$, the poroelastic models are behind the LL model (red) on the main plot and, for $\max(\delta{\sigma^\prime})$, the poroelastic models predict much larger maximum stresses that exceed the vertical scale of the main plot.} \label{fig6-4} }
\end{figure*}

In Figures~\ref{fig6-3} and \ref{fig6-4}, we vary the permeability-load parameter $\zeta$, transitioning the model from impermeable ($\zeta\equiv0$) to fully permeable ($\zeta\equiv1$) for a fixed $\sigma_a=-0.0075$. This value of $\sigma_a$ is between $\sigma_{a}^\mathrm{min}$ and $\sigma_{a}^\mathrm{max}$ for all values of $\zeta$, such that $a<s<b$. Recall that $0\leq\zeta<1$ is analogous to a thin, weak, low-permeability membrane on the inner cavity wall, the permeability of which decreases as $\zeta$ decreases; varying $\zeta$ then varies the partitioning of the fixed total radial stress at the inner cavity wall between fluid loading (injection pressure) and mechanical loading (effective radial stress).

In Figure~\ref{fig6-3}, we plot the same quantities as in Figure~\ref{fig6-1} for four values of $\zeta$ ranging from nearly impermeable ($\zeta\approx0$) to nearly fully permeable ($\zeta\approx1$). For reference, we again plot the true steady-state values of the same quantities (\textit{i.e.}, without subtracting the initial state) in Figure~\ref{zeta_notref} in Appendix~\ref{Figs}.
Note that, because we now drive these models with fixed $\sigma_a$, each model will produce a different value of $q$. We found earlier that increasing nonlinearity led to lower values of $\Delta{p}$ for a given value of $q$ (Figure~\ref{fig6-1}). Here, we then expect that increasing nonlinearity will lead to a higher value of $q$ for a given value of $\sigma_a$, and therefore to larger deformations and stresses. This suggests that the relative ordering of the models should be reversed in Figure~\ref{fig6-3} relative to Figure~\ref{fig6-1}, which is indeed the case.

For the poroelastic models (left column), all quantities increase with $\zeta$. This suggests that, for a given imposed cavity pressure ($-\sigma_a$), a more permeable material will experience larger deformations and larger stresses than a less permeable material. The three quantities $\delta{\phi_f}$, $\delta{\sigma_\theta^\prime}$, and $\delta{p}$ depend quantitatively but not qualitatively on $\zeta$, varying monotonically from a maximum value at the inner radius to a minimum value at the outer radius; the former increases monotonically with $\zeta$, whereas the latter is insensitive to $\zeta$. The two quantities $\delta{u_s}$ and $\delta{\sigma_r^\prime}$ vary both quantitatively and qualitatively with $\zeta$. The disturbance in displacement $\delta{u_s}$ is relatively insensitive to $\zeta$ at the cavity wall, implying that permeability makes very little difference to the size of the cavity. However, $\delta{u_s}$ is increasingly sensitive to $\zeta$ for larger values of $R$, such that the maximum displacement shifts from $R=a_0$ for small $\zeta$ to near $R=1$ for moderate to large $\zeta$. Although the $\delta{u_s}$ curves appear to get closer together as $\zeta$ increases, this is an artefact of the logarithmic vertical scale. The disturbance in radial effective stress $\delta{\sigma_r^\prime}$ vanishes at the outer boundary by construction, and is fixed to $\delta{\sigma_r^\prime}(a)=(1-\zeta)\sigma_a-\sigma_b^\prime$ at the inner boundary. The latter value increases with $\zeta$, transitioning from compressive to tensile, and the rest of the curve essentially pivots about the constant outer boundary value. As a result, $\delta{\sigma_r^\prime}$ is monotonic in $R$ for small $\zeta$, with a minimum at the inner boundary and a maximum at the outer boundary, but transitions to being nonmonotonic in $R$ for large $\zeta$, having an interior maximum and eventually a minimum at the outer boundary.

For the poroelasto-plastic case, the difference between the models grows larger as $\zeta$ increases (right column). This is because plasticity leads to larger deformations, as do larger values of $\zeta$. The plastic region itself also grows larger as $\zeta$ increases (see $\delta{\sigma_\theta^\prime}$). For this value of $\sigma_a$ and sufficiently small $\zeta$, the deformations are small enough that the models are essentially indistinguishable. For this value of $\sigma_a$ and larger values of $\zeta$, the deformations grow sufficiently large that significant differences emerge between the models. For $\delta{\phi_f}$, $\delta{\sigma_\theta^\prime}$, and $\delta{p}$, the consequences of increasing $\zeta$ are again mostly quantitative: The deformations and stresses grow larger. For all models except the LL model, $\delta\sigma_r^\prime$ and $\delta{\sigma_\theta^\prime}$, exhibit weak non-monotonicity in $\zeta$ for some intermediate values of $R$ as $\zeta$ approaches 1. Note that the QL and QQ models predict very large values of $\delta{\phi_f}$ for $\zeta\approx{}1$ relative to the other models and other values of $\zeta$. For $\delta{u_s}$ and $\delta\sigma_r^\prime$, the changes are again more complex. Plasticity leads to much larger displacements at the inner boundary as $\zeta$ increases, but to very little change in the displacement at the outer boundary. The maximum in $\delta{u_s}$ shifts from the inner boundary for all models for small $\zeta$ to the outer boundary for all models for intermediate $\zeta$, and back to the inner boundary for the more nonlinear models for large $\zeta$.

In Figure~\ref{fig6-4}, we plot the same quantities as in Figure~\ref{fig6-2}, but against $\zeta$ for fixed $\sigma_a=-0.0075$. As also illustrated in Figure~\ref{fig6-3}, it is clear that model choice is unimportant for the poroelastic models for all $\zeta$, and for the poroelasto-plastic models for sufficiently small $\zeta$ (for this set of parameters, $\zeta \lesssim 0.55$). Model choice eventually becomes important for the poroelasto-plastic models because plasticity enables large deformations and these deformations increase monotonically with $\zeta$ (Figure~\ref{fig6-4}, top left). As also illustrated in Figure~\ref{fig6-3}, the more nonlinear models again predict larger deformations. The two poroelastic models predict that $\max\left(\delta{\phi_f}\right)$ increases approximately linearly with $\zeta$. These predictions are larger in magnitude than those of the poroelasto-plastic models until $\zeta\approx{}0.55$ (QL and QQ) or $\approx{}0.8$ (LL and NQ), at which point the poroelasto-plastic models increase strongly following a corner where $\max\left(\delta{\phi_f}\right)$ shifts from an internal value to the value at the inner radius. The flow rate $q$ increases monotonically with $\zeta$ for all models, with the more nonlinear models predicting larger flow rates as $\zeta$ approaches 1. Note that $q = -\frac{\zeta\sigma_a}{\ln\left(\frac{b}{a}\right)}$ for all models, and is thus pinned to $q=0$ for an impermeable material ($\zeta\equiv 0$) and to $q = \frac{\Delta p}{\ln\left(\frac{b}{a}\right)}$ for a fully permeable material ($\zeta\equiv 1$). This dependence on $a$ and $b$ links $q$ to the displacement field, and hence to model choice, for $\zeta>0$. Note that, for the L and LL models, $q=\frac{\zeta \sigma_a }{\ln(a^\mathrm{ref})}$ and is therefore independent of deformation. The maximum disturbance in effective stress $\max\left(\delta{\sigma^\prime}\right)$ increases monotonically with $\zeta$ for the L, Q, and LL models, and becomes weakly nonmonotonic in $\zeta$ for the QL, QQ, and NQ models as $\zeta$ approaches 1. This nonmonotonicity is more pronounced for the NQ model. The fact that $\max\left(\delta{\sigma^\prime}\right)$ is maximised for a particular value of $\zeta$ could be important in applications such as hydraulic fracturing in ductile shales or for borehole integrity. Note that the ordering of models for $\max\left(\delta{\sigma^\prime}\right)$ is reversed relative to the ordering of models for $\delta{a}$ and $q$, reinforcing the fact that plasticity dissipates elastic energy and leads to lower (less tensile) stresses. Note, finally, that the poroelasto-plastic models predict significantly larger displacements and significantly lower stresses than the poroelastic models for all values of $\zeta$ (Figure~\ref{fig6-4}, top left and bottom right), whereas all models predict similar flow rates for most values of $\zeta$.

\section{Conclusion}

Fluid-driven deformation is relevant to applications in borehole integrity and cavity expansion; motivated by these problems, this study provides the first kinematically rigorous poroelasto--perfectly-plastic model for fluid injection into a thick-walled annulus. To assess the importance of plasticity, and of large deformations, we performed a detailed examination of four such models: Classical linear poroelasto-plasticity (\textit{i.e.,} linear elasticity with linearised kinematics; LL); linear elasticity with rigorous kinematics in just the plastic region~(QL) and in both regions~(QQ); and linear elasticity, rigorous kinematics, and logarithmic (Hencky) strains in the plastic region~(NQ). For a set of parameter values motivated by sedimentary rocks such as sandstone or shale, we then compared the predictions of these models with each other, and with those of two poroelastic models: Classic linear poroelasticity (\textit{i.e.}, linear elasticity with linearised kinematics; L) and linear elasticity with rigorous kinematics~(Q).

We showed that there was negligible difference between the poroelastic L and Q models for these parameters because the deformations remain small. In contrast, plasticity enables large deformations in the plastic region, making model choice much more important there. Accounting for rigorous kinematics in the plastic region, in particular, can have a significant impact on the predicted behaviour (\textit{e.g.}, compare the LL and QL models). However, this effect is isolated to the plastic region, where deformations are large; in the elastic region, in contrast, deformations remain small and linearised kinematics remain appropriate, even in the presence of large deformations in the plastic region (\textit{e.g.}, compare the QL and QQ models). Linearised kinematics in the plastic region can lead to non-physical predictions (Figure~\ref{figA-unreal} in Appendix~\ref{Figs}).

Previous models have treated low-permeability materials such as shale as either fully permeable or fully impermeable. Here, we proposed a new `permeability-load parameter' $\zeta$ that enables a smooth transition between these limiting states by (essentially) introducing a thin, weak, low-permeability skin at the cavity wall. In Figures~\ref{fig6-1} and \ref{fig6-2}, we considered a fully permeable material ($\zeta\equiv1$) for a range of injection rates $q$. In Figures~\ref{fig6-3} and \ref{fig6-4}, we considered a fixed total stress at the inner radius $\sigma_a$ as $\zeta$ transitions from 0 to 1. We showed that the amount of deformation increases with $\zeta$ for a given value of $\sigma_a$. The maximum tensile effective stress exhibits a maximum at an intermediate value of $\zeta$ near 1, such that an annulus with a slight reduction in permeability at the cavity wall experiences the greatest effective stress. Since the deformation increases with $\zeta$, the choice of poroelasto-plastic model becomes increasingly significant as $\zeta$ increases.

Our results highlight the significant qualitative and quantitative differences between fully impermeable and fully permeable materials, and provide a mechanism for smoothly transitioning between these two end-member behaviours. As such, many practicals scenarios that are currently modelled as either fully impermeable or fully permeable, such as boreholes with clogged or damaged walls, boreholes that have been treated with wall-building chemicals, or boreholes in low-permeability rocks such as shales, are probably best modelled with intermediate values of $\zeta$. We have also shown that, although plastic failure leads to drastically different material behaviour, including much larger deformations and much smaller stresses, it has a relatively minor impact on injection pressure. This indicates that injection pressure is a relatively weak indicator of plastic failure, which may be problematic in practice because pressure is one of the primary observables during injection. In other words, the onset of ductility may be quite difficult to detect from the surface, despite its strong impact on displacements and stresses.

In addition to the above qualitative points about the influence of constitutive behaviour, kinematics, and key parameters, our work here also provides a reference solution that can be used as a rigorous benchmark for finite-element algorithms. It may also be useful for interpreting laboratory experiments in similar geometries~\citep[\textit{e.g.},][]{macminn2015fluid}.

\acknowledgements

The authors are grateful to EPSRC for support in the form of a Doctoral Training Award to L.C.A.

\appendix 

\section{Which yield condition?}\label{which_yield}

We now consider all six possible cohesive Mohr-Coulomb yield criteria for a cylinder in plane strain with principal stresses $\sigma_r^\prime$, $\sigma_\theta^\prime$ and $\sigma_z^\prime$. For plane strain, 
\begin{equation}\label{sigz}
\sigma_z^\prime = \frac{\Gamma}{1+\Gamma}\left(\sigma_r^\prime+\sigma_\theta^\prime\right) \equiv\nu\left(\sigma_r^\prime+\sigma_\theta^\prime\right),
\end{equation}
where $\nu\equiv\frac{\Gamma}{1+\Gamma}$ is the Poisson ratio (\textit{cf.} \S\ref{params}). Recall that $\nu\in(0,\frac{1}{2})$ for most physical materials, and that $\sigma^\prime_1\geq\sigma^\prime_2\geq\sigma^\prime_3$ by definition. We then have three possible constraints based on the signs of $\sigma_r^\prime$ and $\sigma_\theta^\prime$: If $\sigma_r^\prime, \sigma_\theta^\prime>0$, then $\sigma^\prime_1\neq{}\sigma_z^\prime$; if $\sigma^\prime_r,\sigma^\prime_\theta<0$, then $\sigma_3^\prime
\neq\sigma^\prime_z$; and if $\sigma^\prime_r$ and $\sigma^\prime_\theta$ have different signs, or if one of them is zero, then $\sigma^\prime_2\equiv\sigma^\prime_z$. The six possible yield functions are then
\begin{subequations}\label{yield_condns}
    \begin{align}
        &\mathcal{F}_{\theta,r} \defeq \alpha\sigma^\prime_\theta-\sigma^\prime_r-y, \\
        &\mathcal{F}_{\theta,z} \defeq\alpha\sigma^\prime_\theta-\sigma^\prime_z-y, \\
        &\mathcal{F}_{r,\theta} \defeq\alpha\sigma^\prime_r-\sigma^\prime_\theta-y, \\
        &\mathcal{F}_{r,z} \defeq\alpha\sigma^\prime_r-\sigma^\prime_z-y, \\
        &\mathcal{F}_{z,\theta} \defeq\alpha\sigma^\prime_z-\sigma^\prime_\theta-y, \\
        &\mathcal{F}_{z,r} \defeq\alpha\sigma^\prime_z-\sigma^\prime_r-y,
    \end{align}
\end{subequations}
where the material remains elastic for $\mathcal{F}_{1,3}<0$ and yields when $\mathcal{F}_{1,3}=0$. We must then determine which of these yield functions first reaches zero. As stated in \S\ref{sec:plastic}, we assume that yield first occurs at $r=a_\mathrm{min}$; that, prior to yield, the yield function $\mathcal{F}_{1,3}$ is maximised at the poroelastic steady-state and not during the preceding transient evolution; and that, once yield occurs according to a particular yield condition, the material will fail exclusively according to this condition. We are therefore only concerned with the values of $\sigma^\prime_i(a_\mathrm{min})$, where $i\in\{r,\theta,z\}$ (\textit{i.e.}, the effective stresses at the inner boundary at the point of first yield).

\subsection{Fully permeable ($\zeta\equiv1$)}

For $\zeta \equiv 1$, $\sigma^\prime_r(a)=0$ implies that $\sigma_z^\prime(a)=\frac{\Gamma}{1+\Gamma}\sigma^\prime_\theta(a)$ and, as such, that $\sigma^\prime_2\equiv\sigma^\prime_z$. Hence, the only possible yield conditions are $\mathcal{F}_{\theta,r}=0$ and $\mathcal{F}_{r,\theta}=0$. The former is appropriate if $\sigma_\theta^\prime(a) \geq\sigma^\prime_z(a)\geq\sigma^\prime_r(a)=0$, which we assumed throughout our analysis in the main text. We now consider the conditions under which the latter would be appropriate.

\subsubsection{$\mathcal{F}_{r,\theta}=0$}

For $\mathcal{F}_{r,\theta}=0$ to be the appropriate yield condition, it must be the case that $\sigma^\prime_\theta(a)\leq\sigma_z^\prime(a)\leq\sigma^\prime_r(a)=0$. For the L and Q models with $\zeta=1$, it is straightforward to show that (see \S\ref{minsig})
\begin{equation}
    \sigma_\theta^\prime(a) = \frac{2B_2}{a^2}+\frac{q(1-\Gamma)}{2}
\end{equation}
where $B_2$ is given by
\begin{equation}
    B_2(a,b) = \frac{b^2a^2\left[2\sigma_b^\prime+q(1+\Gamma)\ln\left(\frac{b}{a}\right)\right]}{2(b^2-a^2)}.
\end{equation}
We then rearrange the condition $\mathcal{F}_{r,\theta}<0$, for which deformation remains elastic, to give
\begin{equation}
q> q_\star\defeq\frac{2[-(b^2-a^2)y-2b^2\sigma^\prime_b]}{2b^2(1+\Gamma)\ln\left(\frac{b}{a}\right)+(b^2-a^2)(1-\Gamma)}.
\end{equation}
This means that the material will only yield if the injection rate is small enough, or sufficiently negative (suction), to trigger cavity collapse. In the following, we refer to this as `negative yield' and the converse as `positive yield'.

We are only concerned here with fluid injection problems, so we only consider $q\geq0$ (no suction/extraction). As a result, ensuring that $q_\star<0$ would then imply that $\mathcal{F}_{r,\theta}<0$ for $q\geq0$. Rearranging the requirement that $q_\star<0$ leads to the constraint that negative yield cannot occur for all relevant values of $q$ if
\begin{equation}\label{paracrit}
(b^2-a^2)y>-2b^2\sigma_b^\prime,
\end{equation}
meaning that the cylinder must be sufficiently `strong' relative to the compressive far-field stress $\sigma_b^\prime$. This constraint is satisfied for the parameter values used here (see \S\ref{params}). Hence, we can state conclusively that, for a sufficiently strong and fully permeable cylinder, $\mathcal{F}_{\theta,r}$ is the correct yield condition during fluid injection ($q\geq0$).

\subsection{Partially permeable materials ($\zeta\not\equiv1$)}

For $\zeta\in[0,1)$, the radial effective stress at the inner radius is $\sigma_r^\prime(a) = (1-\zeta)\sigma_a\in[-|\sigma_a|,0)$. Hence, we are left with two cases to consider:
\begin{enumerate}
    \item $\sigma^\prime_r(a)<0$ and $\sigma_\theta^\prime(a)>0$, in which case  $\sigma^\prime_2\equiv\sigma^\prime_z$, and $\mathcal{F}_{\theta,r}=0$ and $\mathcal{F}_{r,\theta}=0$ are the possible yield criteria.
    \item $\sigma^\prime_r(a),\sigma_\theta^\prime(a)<0$, in which case $\mathcal{F}_{\theta,r}=0$, $\mathcal{F}_{r,\theta}=0$, $\mathcal{F}_{z,\theta}=0$ and $\mathcal{F}_{z,r}=0$ are the possible yield criteria.
\end{enumerate}
Note that, for $\sigma_r^\prime(a)<0$, the axial effective stress $\sigma_z^\prime(a)$ can never be the minimum principal stress (see discussion above Eqs.~\ref{yield_condns}). Thus, we only have three alternative yield conditions to consider.

\subsubsection{$\mathcal{F}_{r,\theta}=0$}

Ensuring that $\mathcal{F}_{r,\theta}$ is always strictly negative will prevent yield according to $\mathcal{F}_{r,\theta}$. Using Equation~\eqref{genundzeta} leads to
\begin{equation}
    \mathcal{F}_{r,\theta} =(\alpha-1)(1-\zeta)\sigma_a-\left[\frac{2\mathcal{B}_2}{a^2}-\frac{\zeta\sigma_a(1-\Gamma)}{2\ln\left(\frac{b}{a}\right)}\right]-y<0,
\end{equation}
where $\mathcal{B}_2$ is defined in Equation~\eqref{B2apzeta}. Following the same procedure that leads to Equation~\eqref{sigmin}, we find that yield according to $\mathcal{F}_{r,\theta}$ will not occur for
\begin{multline}\label{sigmin3}
    \sigma_a<\sigma_{a;r,\theta}^{\mathrm{min}} = \displaystyle\frac{
2\ln\left(\displaystyle\frac{b }{a }\right)\left[-y(b ^2-a ^2)-2 b ^2\sigma_b^\prime\right]
}{-(\alpha-1)(1-\zeta)+\left\{\zeta(1-\Gamma)\left[2b ^2\displaystyle\ln\left(\frac{b }{a }\right)-b ^2+a ^2\right]-4b ^2\displaystyle\ln\left(\frac{b }{a }\right)\right\}
},
\end{multline}
meaning that $\sigma_{a;r,\theta}^\mathrm{min}$ is the least compressive value of $\sigma_a$ that would prevent negative yield according to $\mathcal{F}_{r,\theta}$. Note that the $\sigma_{a}^\mathrm{min}$ presented in Equation~\eqref{sigmin} is $\sigma_{a;\theta,r}^\mathrm{min}$, corresponding to positive yield according to $\mathcal{F}_{\theta,r}$.

We focus on injection, so it must be the case that $\sigma_a<0$. As a result, ensuring that $\sigma_{a}^{\mathrm{min},r,\theta}>0$ would then imply that $\mathcal{F}_{r,\theta}<0$ for all relevant $\sigma_a$. This constraint simply requires that Equation~\eqref{paracrit} must be satisfied since the denominator of Equation~\eqref{sigmin3} is negative.

\subsubsection{$\mathcal{F}_{z,\theta}=0$}

Ensuring that $\mathcal{F}_{z,\theta}$ is always strictly negative will prevent yield according to $\mathcal{F}_{z,\theta}$. This is the appropriate yield function if $\sigma_\theta^\prime(a)<\sigma_r^\prime(a)<\sigma_z^\prime(a)<0$. If $\alpha\nu\geq1$, as is the case for the parameters used above (\textit{cf.} \S\ref{params}), then $\mathcal{F}_{z,\theta}<0$ rearranges to
\begin{equation}
    \underbrace{\alpha\nu\sigma_r^\prime}_{<0}+\underbrace{\sigma_\theta^\prime(\alpha\nu-1)}_{<0}< \underbrace{y}_{>0},
\end{equation}
which is always satisfied. If $\alpha\nu<1$, the requirement is more complicated. As stated above, this yield condition is only appropriate if $\sigma_\theta^\prime(a)<\sigma_r^\prime(a)$, and therefore cannot be satisfied if $\sigma_r^\prime(a)<\sigma_\theta^\prime(a)$. Substituting Equation~\eqref{thetastres} into the constraint that $\sigma_r^\prime(a)<\sigma_\theta^\prime(a)$, we obtain
\begin{equation}
    \sigma_\theta^\prime(a)-\sigma_r^\prime(a) = \frac{2\mathcal{B}_2}{a^2}-\zeta\frac{\sigma_a(1-\Gamma)}{2\ln\left(\frac{b}{a}\right)}>0.
\end{equation}
Substituting for $\mathcal{B}_2$ from Equation~\eqref{B2apzeta} and rearranging, we find that yield according to $\mathcal{F}_{z,\theta}$ cannot occur if
\begin{equation}
    \underbrace{2b^2\ln\left(\frac{b}{a}\right)}_{>0}\left\{\underbrace{\sigma_a[\zeta(1-\Gamma)-2]}_{>0}+\underbrace{2\sigma_b^\prime}_{<0}\right\} - \underbrace{(b^2-a^2)\zeta\sigma_a(1-\Gamma)}_{>0}>0.
\end{equation}
This is true for sufficiently large $|\sigma_a|$,
\begin{equation}
     |\sigma_a|>\frac{4b^2|\sigma_b^\prime|\ln\left(\frac{b}{a}\right)}{\zeta(b^2-a^2)(1-\Gamma) +2b^2\ln\left(\frac{b}{a}\right)[2-\zeta(1-\Gamma)]}.
\end{equation}
The maximum value of the right hand side of the above occurs in the limit $a\to0$, and is given by $\displaystyle\frac{2|\sigma_b^\prime|}{2-\zeta(1-\Gamma)}$. Thus, the above is always satisfied if
\begin{equation}
    |\sigma_a|>\frac{2|\sigma_b^\prime|} {2-\zeta(1-\Gamma)}.
\end{equation}
This inequality is safely satisfied for all $\zeta$ if
\begin{equation}
    |\sigma_a|>\frac{2|\sigma_b^\prime|}{1+\Gamma},
\end{equation}
since $\mathrm{max}(\zeta)=1$.

\subsubsection{$\mathcal{F}_{z,r}=0$}

Ensuring that $\mathcal{F}_{z,r}$ is always strictly negative will prevent yield according to $\mathcal{F}_{z,r}$. This is the appropriate yield function if $\sigma_r^\prime(a)<\sigma_\theta^\prime(a)<\sigma_z^\prime(a)<0$. The constrain that $\mathcal{F}_{z,r}<0$ rearranges to
\begin{equation}
    \alpha\nu\underbrace{\sigma_\theta^\prime(a)}_{<0} +\underbrace{\sigma_r^\prime(a)}_{<0}(\alpha\nu-1)<y,
\end{equation}
which is again always satisfied if $\alpha\nu>1$. For $\alpha\nu<1$, we appeal to the fact that $\mathcal{F}_{z,r}=0$ will only be the appropriate yield condition if  $\sigma_\theta^\prime(a)<\sigma_z^\prime(a)$, and therefore if $\sigma_\theta^\prime(a)<\Gamma\sigma_r^\prime(a)$. This constraint implies that yield according to $\mathcal{F}_{z,r}$ cannot occur if
\begin{equation}
    |\sigma_a|(1-\Gamma)(1-\zeta) <\frac{2\mathcal{B}_2}{a^2}+\frac{|\sigma_a|\zeta(1-\Gamma)}{2\ln\left(\frac{b}{a}\right)}
\end{equation}
which rearranges to 
\begin{equation}
    |\sigma_a|>\frac{4b^2|\sigma_b^\prime| \ln\left(\frac{b}{a}\right)}{2\ln\left(\frac{b}{a}\right)\left[b^2(1+\Gamma)+a^2(1-\Gamma)(1-\zeta)\right]+\zeta(b^2-a^2)(1-\Gamma)}.
\end{equation}
Again, the right-hand side is maximised for $a\to0$ and $\zeta\to{}1$, and is therefore satisfied for all $\zeta$ if
\begin{equation}
    |\sigma_a|>\frac{2|\sigma_b^\prime|}{1+\Gamma},
\end{equation}
which is the same constraint derived above for $\mathcal{F}_{z,\theta}$.

\subsection{Summary}

In summary, for fixed $q$ and $\zeta\equiv1$, there are only two possible yield conditions: $\mathcal{F}_{\theta,r}=0$ and $\mathcal{F}_{r,\theta}=0$. The latter corresponds to collapse or negative yield, and there is a value of $q=q^\star$ above which this cannot occur. We showed above that the cylinder will yield exclusively according to the former for any $q\geq0$, provided that the cylinder is sufficiently `strong' relative to the compressive far-field stress---that is, if Equation~\eqref{paracrit} is satisfied.

For a fixed total stress at the inner radius, $\zeta\in[0,1)$, there are more possibilities; however, if $\alpha\nu>1$ and Equation~\eqref{paracrit} is satisfied, then yield will occur exclusively according to $\mathcal{F}_{\theta,r}$ for all $\sigma_a\leq0$. Note that $\mathcal{F}_{r,\theta}$ and $\mathcal{F}_{z,\theta}$ both model negative yield (cavity collapse). Even if $\alpha\nu<1$, $\mathcal{F}_{\theta,r}=0$ remains the appropriate yield condition for all $\zeta$ and $a$, provided that $|\sigma_a|>\displaystyle\frac{2|\sigma_b^\prime|}{1+\Gamma}$.

\clearpage
\section{Additional Figures}\label{Figs}

\begin{figure}[b]
    \begin{center}
        \includegraphics[width=0.97\textwidth]{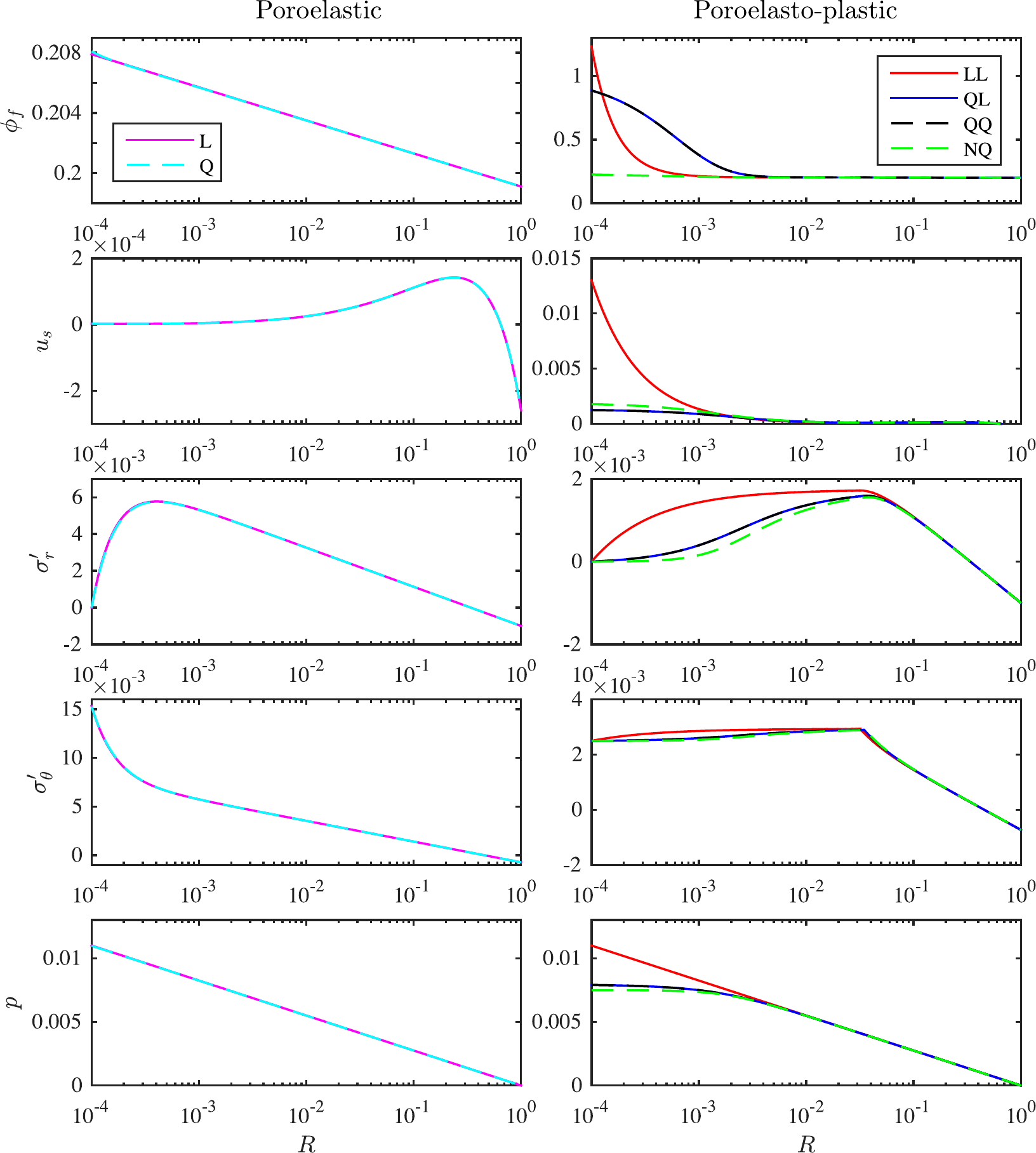}
    \end{center}
    \caption{\small{As Figure~\ref{fig6-1}, but without subtracting the compressed initial state. Note that, unlike in Figure~\ref{fig6-1}, the top two rows are no longer on a logarithmic vertical scale.} \label{figA-noref} }
\end{figure}
\begin{figure}[tbp]
    \begin{center}
    \includegraphics[width=1.0\textwidth]{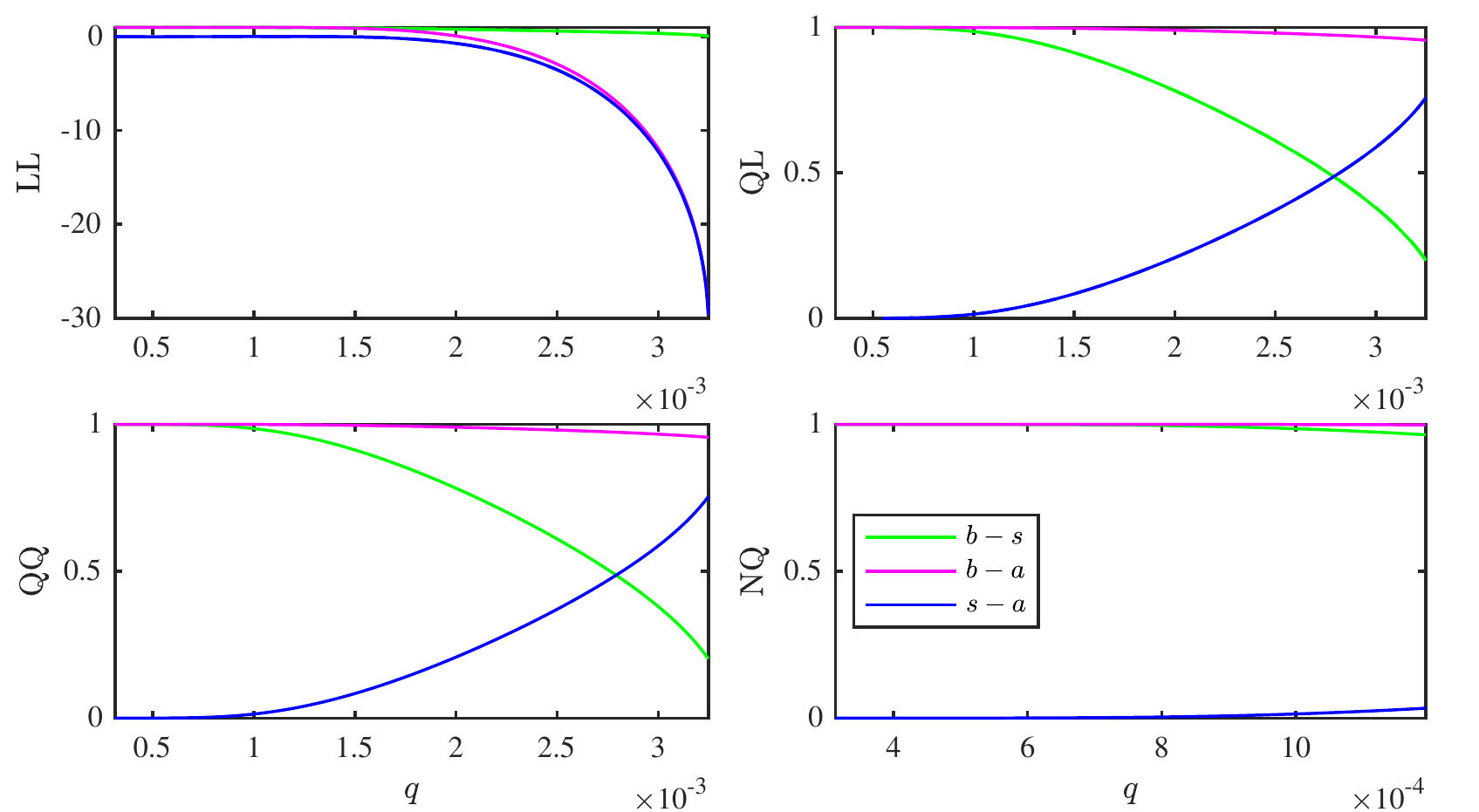}
    \end{center}
    \caption{\small{For all four poroelasto-plastic models---LL (top left), QL (top right), QQ (bottom left), and NQ (bottom right)---we plot the total thickness of the annulus $b-a$ (magenta), the thickness of the elastic region $b-s$ (green), and the thickness of the plastic region $s-a$ (blue) against flow rate $q$ for a fully permeable material ($\zeta\equiv1$). Note that for the LL, QL, and QQ models, $q$ is between $q_\mathrm{min}\approx 3\times10^{-4}$, the smallest $q$ that induces yield in all models, and $q_\mathrm{max}\approx 3.3\times10^{-3}$, the first value of $q$ for which the material is entirely yielded. Note that the range of $q$ is reduced for the NQ model relative to the others. The LL model is the only one of these that linearises the kinematics in the plastic region, where deformations are largest. As a result, the LL model predicts non-physical behaviour for sufficiently large $q$ (\textit{i.e.}, $s<a$ and $b<a$); the other predict physical results for all $q$ until the material has yielded entirely.} \label{figA-unreal} }
\end{figure}
\begin{figure}[tbp]
    \begin{center}
        \includegraphics[width=1.0\textwidth]{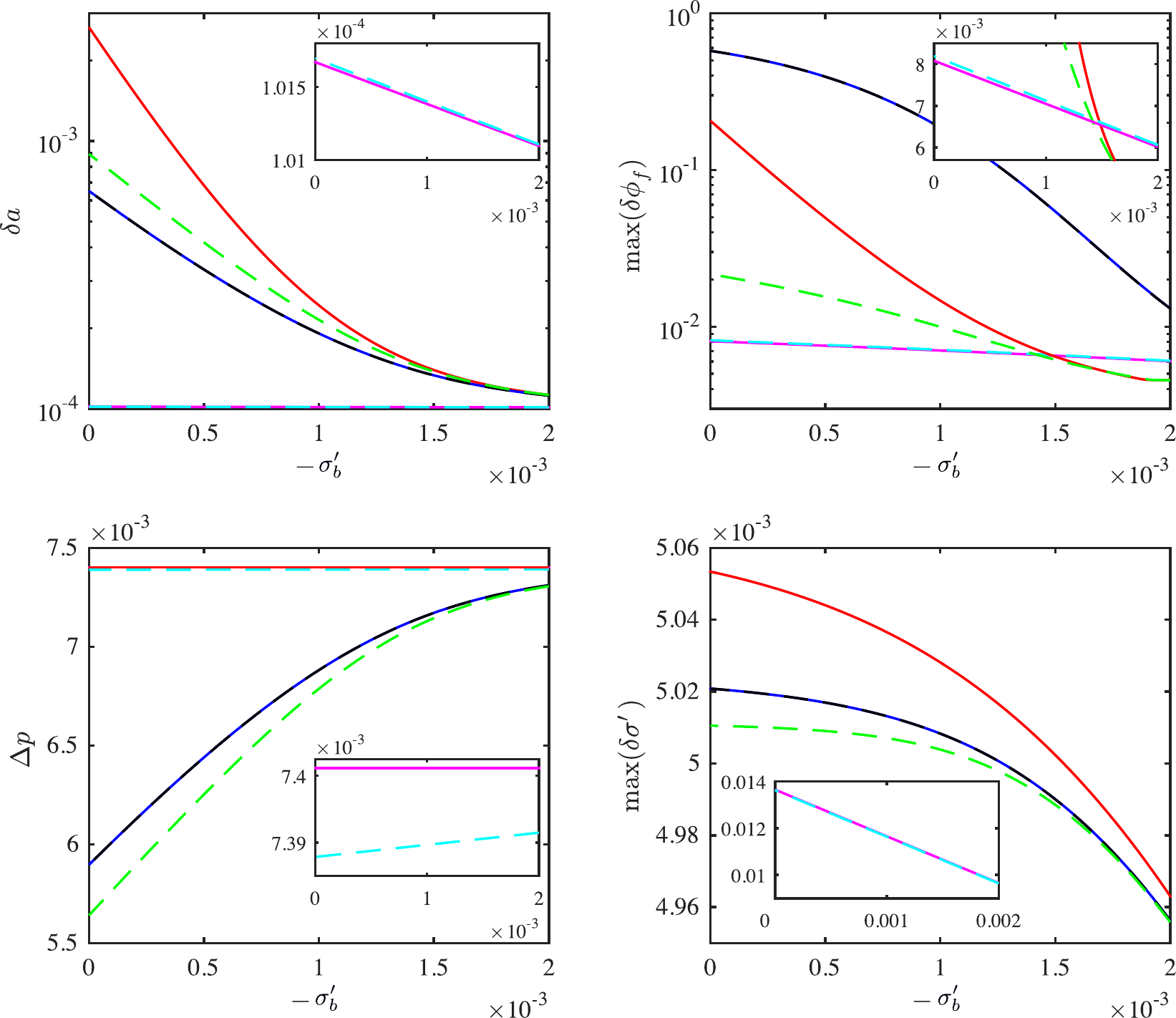}\end{center}
        \caption{\small{As Figures~\ref{fig6-2} and \ref{fig6-4}, but plotted against varying $-\sigma_b^\prime$ for a fully permeable material ($\zeta\equiv1$). This illustrates the transition from an unconstrained cylinder ($\sigma_b^\prime\equiv0$) to a highly constrained cylinder, providing a link to our previous work on fully permeable unconstrained poroelastic cylinders \cite{auton2017arteries, auton2018arteries}.} \label{figA-sig} }
\end{figure}
\begin{figure}[tbp]
    \vspace{-1cm}
    \begin{center}
        \includegraphics[width=1.0\textwidth]{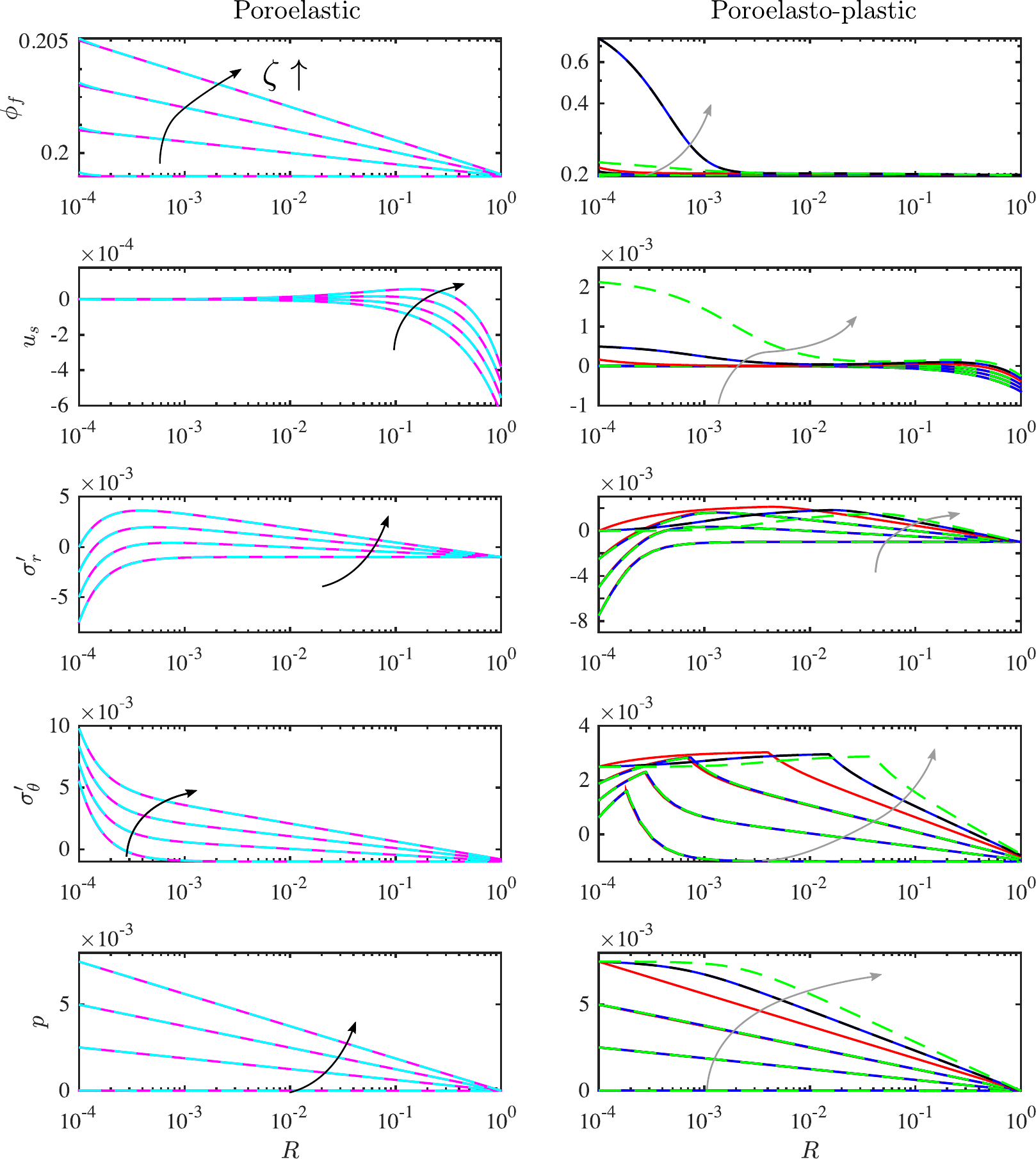}
    \end{center}
    \caption{\small{As Figure~\ref{fig6-3}, but without subtracting the compressed initial state. Note that the top right panel is now on a logarithmic vertical scale to show the behaviour near the inner boundary for all values of $\zeta$.} \label{zeta_notref} }
\end{figure}

\clearpage

\bibliographystyle{plainnat} 


\end{document}